\newcommand{\CLASSINPUTtoptextmargin}{0.85in}
\newcommand{\CLASSINPUTbottomtextmargin}{1in}
\documentclass[journal]{IEEEtran}
\makeatletter
\def\ps@headings{%
	\def\@oddhead{\mbox{}\scriptsize\rightmark \hfil \thepage}%
	\def\@evenhead{\scriptsize\thepage \hfil \leftmark\mbox{}}%
	\def\@oddfoot{}%
	\def\@evenfoot{}}
\makeatother
\IEEEoverridecommandlockouts

\usepackage{setspace}
\usepackage{amsmath}
\usepackage{amsfonts}
\usepackage{mathrsfs}
\usepackage{amssymb}
\usepackage{cases}
\usepackage{stmaryrd}
\usepackage{mathrsfs}
\usepackage{bbm}
\usepackage[dvips]{graphicx}
\usepackage{algorithm}
\usepackage{multirow}
\usepackage{graphicx}
\usepackage{subfigure}
 \usepackage{cite}
\usepackage{color}
\usepackage[table]{xcolor}

\newtheorem{lm}{Lemma}
\newtheorem{Th}{Theorem}
\newtheorem{propRemark}{Remark}
\newtheorem{corollary}{Corollary}
\newtheorem{Definition}{Definition}

\definecolor{chao}{rgb}{0,0,1}
\begin{document}
	

\title{Physical Layer Security Enhancement Using Artificial Noise in Cellular Vehicle-to-Everything (C-V2X) Networks}
%
%
%

\author{Chao~Wang,~
        Zan~Li,~\IEEEmembership{Senior Member, IEEE},
        Xiang-Gen Xia,~\IEEEmembership{Fellow, IEEE},
        Jia~Shi,~\IEEEmembership{Member, IEEE},
         Jiangbo~Si,~\IEEEmembership{Member, IEEE}
        and Yulong Zou,~\IEEEmembership{Senior Member, IEEE}

\thanks{C. Wang, Z. Li, J. Shi, and J. Si are  with the Integrated Service Networks Lab,
Xidian University, Xi'an 710071, China (e-mail: drchaowang@126.com).
}
\thanks{X.-G. Xia is with the College of Communications Engineering, Xidian
	University, Xi'an 710071, China, and also with the Department of Electrical and
	Computer Engineering, University of Delaware, Newark, DE 19716, USA(e-mail: xianggen@udel.edu). }
\thanks{
Y. Zou is with the School of Telecommunications
and Information Engineering, Nanjing University of Posts and Telecommunications, Nanjing 210003, China (e-mail: yulong.zou@njupt.edu.cn).
}
}

%
%


\maketitle

\begin{abstract}
The secure transmission of confidential information in  cellular vehicle-to-everything (C-V2X) communication networks is vitally important for user's personal safety. However, for C-V2X there  have not been much studies on  the physical layer security (PLS). Since  artificial noise (AN) and secure beamforming are popular PLS techniques for  cellular communications, in this paper we investigate the potential of these PLS techniques  for enhancing the security of  C-V2X networks. In particular,
leveraging  stochastic geometry, we study the PLS of an AN  assisted C-V2X network, where the locations of legitimate vehicular nodes, malicious vehicular nodes  and road side units (RSUs) are modeled by Cox processes driven by a common Poisson line process (PLP), and the locations of  cellular base stations (BSs) are modeled by a two-dimensional (2D) Poisson point process (PPP).
Based on the maximum signal-to-interference-ratio (SIR) association scheme, we  calculate the coverage probability of the network.
We also derive bounds on the secrecy probability,  which are validated by simulation results. Moreover, we  obtain an analytical result of the effective secrecy throughput for characterizing the reliability and security of  wiretap channels.
Simulation results are given to validate the analytical result, and provide interesting  insights into
the impact of network parameters on the achievable secrecy performance. Simulation results show that a larger array antenna can provide a better  robustness of the secure transmission strategy, and the optimal power allocation ratio between  confidential information and AN remains almost unchanged  for different numbers of antennas.
\end{abstract}


\IEEEpeerreviewmaketitle

\section{Introduction}
Vehicle-to-everything (V2X) communication, including vehicle-to-vehicle (V2V), vehicle-to-infrastructure (V2I), vehicle-to-network (V2N), and  vehicle-to-pedestrians (V2P), is a key technology for pushing intelligent transportation systems (ITS) forward and improving  road traffic safety. In general, there are two key radio access technologies (RATs) for V2X, including dedicated short range communications (DSRC) and cellular-V2X (C-V2X).
It is known that, C-V2X has been developed by 3GPP in its Rel. 14 \cite{LTEV2X}, operating in the cellular operators' licensed spectrum under the existing LTE  system architectures.
Nevertheless, C-V2X communications have faced huge secrecy challenges, i.e.,  not only incurring the confidential information leakage, but also \textbf{endangering the personal safety of  users, since the information about identifications, positions, and trajectories of users could be exposed to track vehicles easier  \cite{CV2XSecurity}}.
However, as pointed out by \cite{CV2XSecurity}, 3GPP specifications provide few mechanisms for securing  C-V2X communications.
Thus, more research efforts should be devoted to a new C-V2X security arrangement \cite{SecureV2X}.
\subsection{Related Works}
For  C-V2X communications,
there are two widely-used secure transmission strategies \cite{ZouSecurity}: key-based cryptography technologies \cite{SurveyCryptographic} and physical layer security (PLS) techniques \cite{PLS5G,CV2XPLS,VehicleCommunicationSecrecyCommunication,PLSIntelligentCVN}.  
Key-based cryptography technologies guarantee communication security at  upper layers, which introduce much communication overhead due to the key distribution. For asymmetric cryptographic algorithms, the computational and network resources for carrying out the key negotiation are  high. 
These issues make the key management of cryptography technologies become more challenging for  delay-sensitive  C-V2X applications \cite{CV2XPLS}.
In addition, with the development of  powerful computers, especially quantum computers, traditional cryptography technologies will be vulnerable to brute-force attacks in the near future, which may lead to security vulnerability in future C-V2X systems \cite{CybersecurityQuantum}. Different from cryptography technologies, PLS technologies  exploit characteristics of wireless medium to design transmission strategies at the physical layer for improving the communication security \cite{VehicleCommunicationSecrecyCommunication}, which do not rely on the computational complexity and do not need trusted authority and tamper-proof devices \cite{PLSIntelligentCVN,PLS5G}.
Therefore, compared with cryptography technologies, the implementation complexity and network overhead of PLS technologies are both much lower, which makes the PLS technology be an effective measure for improving  the security of  delay-sensitive C-V2X applications \cite{PLSIntelligentCVN,SecrecyRateMaximizationV2X}. 
A comprehensive
overview on PLS assisted vehicle networks was given in \cite{PLSIntelligentCVN}.
The secrecy rate maximization of underlaying V2V communications has been studied in \cite{SecrecyRateMaximizationV2X}.
But, until now, there has been no study investigating the aggregate network performance of  C-V2X secure communications \cite{PhysicalLayerSecuritySurvey}.

Generally, there are two popular approaches for investigating the aggregate network performance, including system-level simulations and stochastic geometry. In \cite{Vehicle3GPP}, a system-level simulation based approach has been adopted for studying the performance of C-V2X, which, however, is time-consuming and lacks of analytical  expressions. In contrast,  stochastic geometry based approaches can give an elegant analytical result, thereby {facilitating} the analysis of the effect of network parameters on the network performance. Thus, it  has been widely used for evaluating the  performance of a heterogeneous cellular network \cite{StochasticGeometryRandomGraphs}.
Recently, the stochastic geometry-based approach for evaluating the network performance of C-V2X communications is gaining the attention. For example,  in \cite{PoissonCoxProcess}, the Poisson line process (PLP) was used to model the spatial distributions of roads, and a one-dimensional (1D) Poisson point process
(PPP) was adopted for modeling the spatial distribution of the network nodes on each road, e.g., vehicular nodes and  road side units (RSUs).
A tractable framework for characterizing the downlink coverage performance of urban millimeter wave
vehicular networks has been given in \cite{V2Icommunications}.
Employing the Poisson Cox point process model, a tractable analytical
framework for analyzing the downlink communication of cellular networks leveraging vehicles has been given in \cite{AnalyticalFrameworkVehicleNetwork}.  


All of the works mentioned above have not addressed the security of  C-V2X networks.
{Using} stochastic geometry, the PLS of  wireless communications has been widely studied from the viewpoint  of the network, but most of the works focused on cellular communications. 
Moreover, the PLS of the artificial noise (AN) aided ad hoc network has been studied in  \cite{EnhancingSecrecyMultiAntenna}, where  the secrecy throughput of   sectoring and beamforming strategies was investigated.  
The work \cite{PhysicalLayerSecurityMillimeter} has analyzed the secure connectivity probability and secrecy throughput of the AN-assisted millimeter wave network. 

\subsection{Motivations and Contributions}
 To  our knowledge, there are no published
references that study the PLS of  multi-antenna C-V2X communications from the  viewpoint of the network.
\textbf{In this paper, we aim to provide  an analytical framework to study the PLS of  multi-antenna C-V2X networks using the stochastic geometry approach. It is  to study the potential of the AN aided multi-antenna secure transmission strategy for improving the PLS of C-V2X networks.} Our contributions are summarized as follows.

1) {We build an analytical model for analyzing the PLS of the multi-antenna C-V2X network}, where the spatial distributions of  legitimate vehicular nodes,  RSUs and  malicious vehicular eavesdroppers (Eves)  on  roads are modeled by  Cox processes with a common PLP, and the spatial distributions of  cellular base stations (BSs) and cellular users are modeled by independent two-dimensional (2D) PPPs. With such models, we study the potential of the multi-antenna technique for enhancing the PLS of  C-V2X networks, where AN is adopted for disturbing  randomly located Eves.

2) {We analyze the coverage probability of the multi-antenna C-V2X network.}  Different from  cellular networks,  the characteristics of  future C-V2X services can be summarized as  extremely high data rates,  high reliability, and  low latency, due to the requirements of self-driving autonomous cars. Therefore, 
the max signal-to-interference-ratio (SIR) association policy is adopted in this paper. Although the exact analytical result is difficult to obtain due to the coupling interference at the receiver, an approximate analytical result of the coverage probability is obtained by employing the bound on the distribution of  gamma random variable. Furthermore, asymptotic analytical results show that the Laplace transform of the interference power converges to a 2D PPP model. 

3) {We analyze the secrecy probability of the multi-antenna C-V2X network}, which is defined as the probability that the maximum SIR of  multiple Eves is below a threshold. Although it is challenging to obtain the exact analytical result, we derive   lower and upper bounds on the secrecy probability.
In addition, asymptotic analytical results show that the probability distribution function (pdf) of the minimum distance from multiple Eves to the typical transmitter converges to a 2D PPP model.
Simulation results show that the lower bound  is very tight. Meanwhile, we introduce a secrecy performance metric, namely \textbf{effective secrecy throughput for quantifying the average data rate of the confidential information  that is securely transmitted}, which characterizes the tradeoff between  reliable transmission (coverage probability) and secure transmission (secrecy probability).

4) {We evaluate the impact of the key  parameters on the effective secrecy throughput of the network}, including the number of transmit antennas, the power allocation coefficient between AN and confidential signals, and the intensity of network nodes, etc. We observe that  increasing the number of antennas not only improves the effective secrecy throughput, but also increases the robustness of the AN-assisted secure transmission strategy.
Simulation results show that the effective secrecy throughput may be a concave/quasi-concave function of the  power allocation ratio, and the optimal power allocation ratio remains almost unchanged  with different  numbers of antennas. 

\emph{Notation:} $x\sim\textrm{gamma}(k,m)$ denotes the gamma-distributed random variable with shape $k$ and scale $m$. $x\sim\textrm{exp}(b)$ denotes that $x$ is an exponential random variable whose
mean is $b$. $\mathbf{x} \sim \mathcal{CN}\left(\mathbf{\Lambda}, \mathbf{\Delta}\right)$ denotes  the circularly symmetric complex Gaussian vector with the mean vector $\mathbf{\Lambda}$ and covariance matrix $\mathbf{\Delta}$.
 The factorial of a non-negative integer $n$ is denoted by $n!$. $\Gamma(x)$ is the gamma function.  $\binom{n}{k}=\frac{n!}{\left(n-k\right)!k!}$.
 $||\cdot||_F$ denotes the Frobenius norm. 
 $\mathcal{L}_X(s)$ denotes the Laplace transform of $X$, i.e., $\mathbb{E}\left(e^{-sX}\right)$, $\mathbb{C}^{n\times n}$ stands for a $n\times n$ complex matrix.
 Besides the  notations mentioned above, some important notations are summarized
 in Table I.

\begin{table}[!hbp]
	\centering
	\caption{List of some important notations}
	\begin{tabular}{| l | p{6cm} |}
		\hline
		Symbols & Description \\ \hline
		${\Phi}_b,{\Phi}_u,{\Phi}_e$ & Planar transmitters, planar receivers, and planar Eves  \\
	    ${\Psi}_b,{\Psi}_u,{\Psi}_e$ & Vehicular transmitters, vehicular receivers, and vehicular Eves  \\
	    ${\Phi}_l$ & Poisson line process \\
		${\psi}_b(l)$ & Vehicular transmitters  on the road $l$ \\	
	    ${\psi}_u(l)$ & Vehicular receivers  on the road $l$ \\
	    ${\psi}_e(l)$ & Vehicular eavesdroppers  on the road $l$ \\					
		$\lambda_b, \lambda_u, \lambda_e,\lambda_l$ & Intensities of ${\Phi}_b, {\Phi}_u, {\Phi}_e,{\Phi}_l$ \\
		$u_b,u_u,u_e$ & Intensities of ${\psi}_b(l), {\psi}_u(l), {\psi}_e(l)$\\
		$\phi$ & Power allocation ratio between confidential information and artificial noise.\\
		 \hline
	\end{tabular}
\end{table}

\section{System Model}
We consider the multi-antenna C-V2X communication network, which  consists of vehicular
nodes, RSUs, macro base stations (MBSs), cellular users, and  passive Eves. In the network, there are different types of transmissions, including V2V, V2N, V2P, and V2I, etc.  Without loss of generality, malicious vehicular nodes or malicious cellular users could play as Eves for wiretapping the privacy-sensitive information.
As we know, for C-V2X communications, the vehicular safety applications
require high-reliable and high-capacity communications. In addition,
with the increasing amount of automation in vehicles, the requirements of the C-V2X communications become increasingly high. Therefore, compared with the cellular communications, the C-V2X communications need higher wireless link quality \cite{CV2XSecurity}, which is dominated by the received SIR.
Therefore, we employ the max-SIR association policy suggested by  \cite{KtieHeterogeneous}, for characterizing the maximum secrecy performance, where the user is associated with the BS that provides the maximum SIR.
Note that, we only consider the case that the SIR threshold above 0 dB for studying the coverage probability in order to meet the requirements of vehicular communications and we ignore other trivial cases.

\subsection{Spatial Modeling}
\begin{figure}[!t]
	\centering
	\includegraphics[width=2.5in]{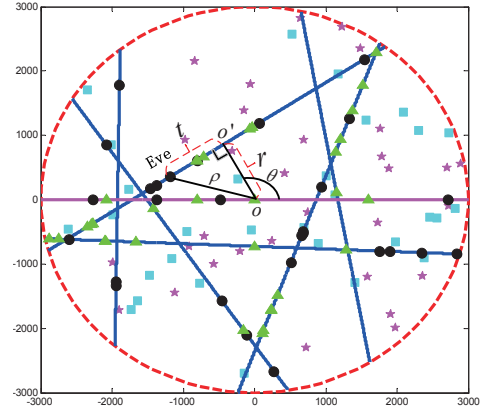}
	\caption{Stochastic geometry-based model of the C-V2X secure transmission in a circular simulation window with radius 3km, road (blue line), vehicular transmitters (green triangles), planar transmitters (cyan squares), planar Eves (pink pentagrams), and vehicular Eves (black squares) with intensity $\lambda_l=10^{-3}/\mathrm{m}$, $\lambda_b=10^{-6}/\mathrm{m}^{2}$,$\lambda_e=10^{-6}/\mathrm{m}^{2}$,
		$u_b=10^{-3}/\mathrm{m}$,$u_e=10^{-3}/\mathrm{m}$. }
	\label{NetworkTopology}
\end{figure}
Fig. \ref{NetworkTopology} gives a realization of the stochastic geometry model for  secure transmissions in  C-V2X networks. We assume that each transmitter is equipped with $N$ antennas, and all of the receivers and Eves are all equipped with a single antenna.
All the cellular MBSs, legitimate cellular users and malicious cellular users are termed as\textbf{ planar transmitters}, \textbf{planar receivers}, and \textbf{planar Eves}, which are, respectively, modeled by three independent PPPs, $\Phi_b$, $\Phi_u$, $\Phi_e$ with intensities $\lambda_b$, $\lambda_u$, $\lambda_e$. Further, all the legitimate vehicular nodes, malicious vehicular nodes, and RSUs  are modeled by  Cox point processes driven by {a common PLP}. The spatial distribution of roads is modeled by a motion-invariant PLP $\Phi_l$ with line intensity $\lambda_l$, which is produced by a 2D PPP $\Xi_l$ with intensity $u_l=\frac{\lambda_l}{\pi}$, on the representation space.
The mathematical preliminary about the PLP can be found in \cite{PoissonCoxProcess,AnalyticalFrameworkVehicleNetwork}.
The road $l_{r,\theta}$ is uniquely parameterized by a point $\left(r,\theta\right)$ of $\Xi_l$. As shown by Fig. \ref{NetworkTopology}, $r$ denotes the perpendicular distance of the road from the origin, while $\theta$ denotes the angle between the positive x-axis and the perpendicular line from the origin to the road in
the counter clockwise direction.
In this case,  the distance $\rho$ from the vehicular Eve on the road to the origin is $\rho=\sqrt{r^2+t^2}$.
On the other hand, all the transmitters, including vehicular nodes and RSUs, on each road $l\in\Phi_l$ are termed as\textbf{ vehicular transmitters},
which are modeled by a homogeneous 1D PPPs $\psi_b(l)$ with the intensity $u_b$.
 All the legitimate and malicious receivers, including vehicular nodes and RSUs, on each road are termed as \textbf{vehicular receivers} and \textbf{vehicular Eves}, which are modeled by  independent homogeneous 1D PPPs $\psi_u(l)$, $\psi_e(l)$ with intensities $u_u$, $u_e$, respectively.
 We denote the set of the vehicular Eves, vehicular users, and vehicular transmitters by $\Psi_e=\left\{\psi_e(l)\right\}_{l\in\Phi_l}$,  $\Psi_u=\left\{\psi_u(l)\right\}_{l\in\Phi_l}$, and $\Psi_b=\left\{\psi_b(l)\right\}_{l\in\Phi_l}$, respectively.


\subsection{AN Assisted Secure Transmission}
Employing the PLS technique, each transmitter sends AN along with the confidential information for confusing the potential Eves \cite{ArtificialNoiseII}. Assume that the perfect CSI of the intended receiver is available at each transmitter, which adopts the maximum ratio transmission (MRT) strategy for transmitting the confidential information while sending AN uniformly in the null-space of the intended channel \cite{EnhancingSecrecyMultiAntenna}.

Specially, let us denote the intended channel of the transmitter at $x\in\mathbb{R}^2$ as $\mathbf{f}_x\in\mathbb{C}^{N\times 1}$ which is a complex Gaussian random vector obeying $\mathcal{CN}\left(\mathbf{0},\mathbf{I}_N\right)$. Its null-space is $\mathbf{W}_x\in\mathbb{C}^{N\times (N-1)}$. Hence, the signal vector $\mathbf{y}_x$ transmitted from the transmitter located at $x$ is 
\begin{align}
\mathbf{y}_x = \sqrt{\phi P_t}\frac{\mathbf{f}_x}{||\mathbf{f}_x||_F}s
+\sqrt{\frac{(1-\phi)P_t}{\left(N-1\right)}}\mathbf{W}_x \mathbf{n}_a,
\end{align}
where $s$ is the confidential signal. In addition, we have the following notations: $\mathbf{n}_a\in\mathbb{C}^{(N-1)\times 1}$ is the AN vector, which is a complex Gaussian random vector obeying $\mathcal{CN}\left(\mathbf{0},\mathbf{I}_{N-1}\right)$, $P_t$ is the total transmit power at the transmitter, $\phi$ is the power allocation ratio. In this paper, we consider the interference-limited scenario, where the thermal
noise is ignored due to  the aggregate interference, which is a common assumption adopted by the existing works on the cellular network analysis using stochastic geometry  \cite{EnhancingSecrecyMultiAntenna}.
For brevity, we set $P_t=1$.
\subsection{Wiretap Coding and Secrecy Performance Metric}
Prior to the transmission, the confidential information is encoded with Wyner code \cite{Wyner}, and we adopt the \textbf{nonadaptive encoder} with the fixed codeword rate $R_b$, and the confidential information rate $R_s$ \cite{RethinkingSecrecy}. Hence, the rate redundancy $R_e\triangleq R_b-R_s$ is added intentionally for protecting the confidential information against malicious eavesdropping. 
We focus on the slow-fading scenario, where the  coherence time of the channel is much longer than the symbol duration\footnote{{Such assumption holds in low-mobility scenarios, such as C-V2X communications in an urban setting}}.  
Suggested by \cite{EnhancingSecrecyMultiAntenna,PhysicalLayerSecurityMillimeter,OptimizationCodeRate}, we employ a probabilistic secrecy performance measure, namely \textbf{the effective secrecy throughput}, which quantifies the average confidential data rate transmitted securely from each transmitter to its intended user.
Before proceeding, the following  definitions are introduced.
\begin{itemize}
\item \textbf{Coverage probability:}
When the capacity of the legitimate link can support the code rate $R_b$, the coverage can be guaranteed, and the legitimate receiver can decode the signals with negligible errors. The coverage probability is defined as $p_{c}\triangleq\mathrm{Pr}\left(\mathrm{SIR}_u\geq \gamma\right)$, where $\mathrm{SIR}_u$ is the SIR received at the typical user and $R_b=\mathrm{log}\left(1+\gamma\right)$.  Compared with traditional cellular users, C-V2X terminals will demand a higher link quality to ensure the  vehicular safety, such as the control information transmission for automatic driving, etc. \textbf{Hence, we only consider the SIR threshold $\gamma>0$ dB, which fits into most cases of the future C-V2X networks.}
\item \textbf{Secrecy probability:}
When the capacity of the eavesdropping link is below the code rate $R_e$, the security of the confidential information can be guaranteed, and Eve can not decode the confidential information. The secrecy probability is defined as $p_{sec}\triangleq\mathrm{Pr}\left(\max_e\mathrm{SIR}_{e}\leq \beta\right)$, where $\max_e\mathrm{SIR}_{e}$ is the maximal SIR received at multiple Eves and $R_e=\mathrm{log}\left(1+\beta\right)$.
\end{itemize}

We adopt the effective secrecy throughput given in  \cite[Definition 1]{OptimizationCodeRate} for characterizing the network security. For the self-containedness,  its definition
 is given as follows:
 \begin{Definition}
 	The effective secrecy throughput is defined as
\begin{align}
\eta \triangleq R_sp_{c}p_{sec}=\left(\mathrm{log}\left(1+\gamma\right)-\mathrm{log}\left(1+\beta\right)\right)p_{c}p_{sec}.\label{SecrecythroughputDefinition}
\end{align}
\end{Definition}
\subsection{Some Important Laplace Transforms}
In order to carry out the coverage probability and secrecy  probability analysis, one should have the knowledge about the Laplace transforms of the interference power originating from different types of transmitters.
\begin{lm}
Denoting $\zeta_x$ as the equivalent channel gain from the transmitter at $x$ to the  typical receiver at the origin and $\alpha>2$ as the path loss exponent, the Laplace transforms of the interference power received by the typical receiver located at the origin, originating from the \textbf{set of vehicular transmitters} $\Psi_b$, from the \textbf{set of planar transmitters} $\Phi_b$, and from the \textbf{set of the vehicular transmitters on the road at a distance $r$}, i.e., $\psi_b(l_{r})$,
are, respectively, given by
\begin{align}
&\mathcal{L}_{\Psi_b}(s)\triangleq\mathbb{E}\left(\mathrm{exp}\left(-\sum_{x\in\Psi_b}{s\zeta_xD_{xo}^{-\alpha}}\right)\right)
\nonumber\\
&{=}\mathrm{e}^{-2\lambda_l\int^{+\infty}_0\left(1-e^{-2u_b\int^{+\infty}_0\left(1-\mathcal{L}_{\zeta_x}\left(s\left(r_b^2+t_b^2\right)^{-\frac{\alpha}{2}}\right)\right)dt_b}\right)dr_b},
\label{LaplacePSibLemma}
\end{align}
\begin{align}
&\mathcal{L}_{\Phi_b}(s)\triangleq\mathbb{E}\left(\mathrm{exp}\left(-\sum_{x\in\Phi_b}{s\zeta_xD_{xo}^{-\alpha}}\right)\right)
\nonumber\\
&{=}\mathrm{exp}\left(-\lambda_b\pi\mathbb{E}\left(\zeta_x^{\delta}\right)\Gamma(1-\delta)s^{\delta}\right),
\label{LaplacePhibLemma}\\
&\mathcal{L}_{\psi(l_{r})}(s)\!\triangleq\!\mathbb{E}\left(\mathrm{exp}\left(-\sum_{x\in\psi_b(l_{r})}{s\zeta_x}{D_{xe}^{-\alpha}}\right)\right)
\nonumber\\
&\!\!=\!\!\mathrm{exp}\left(-2u_b\int^{+\infty}_0\left(1-{\mathcal{L}_{\zeta_x}\left(s\left(r^2+t_b^2\right)^{-\frac{\alpha}{2}}\right)}\right)dt_b\right).
\label{LcpsilrLemma}
\end{align}
\end{lm}
\begin{IEEEproof}
The proof of the Laplace transform (\ref{LaplacePSibLemma}) is given as
\begin{align}
&\mathbb{E}\left(\mathrm{exp}\left(-\sum_{x\in\Psi_b}{s\zeta_xD_{xo}^{-\alpha}}\right)\right)
\nonumber\\
&{=}\mathbb{E}\left(\prod_{(r_b,\theta_b)\in\Xi_{l}}\mathbb{E}\left(\prod_{t_b\in\psi(l_{r,\theta})}\mathrm{exp}\left(-{s\zeta_x}{\left({r_b^2+t_b^2}\right)^{-\frac{\alpha}{2}}}\right)\right)\right)
\nonumber\\
&\overset{(a)}{=}\mathrm{e}^{-2\lambda_l\int^{+\infty}_0\left(1-e^{-2u_b\int^{+\infty}_0\left(1-\mathrm{E}_{\zeta_x}\left(\mathrm{exp}\left(-s\zeta_x\left(r_b^2+t_b^2\right)^{-\frac{\alpha}{2}}\right)\right)\right)dt_b}\right)dr_b}
\nonumber\\
&{=}\mathrm{e}^{-2\lambda_l\int^{+\infty}_0\left(1-e^{-2u_b\int^{+\infty}_0\left(1-{\mathcal{L}_{\zeta_x}\left(s\left(r_b^2+t_b^2\right)^{-\frac{\alpha}{2}}\right)}\right)dt_b}\right)dr_b},
\end{align}
where, step $(a)$ is obtained by using the PGFL of $\psi(l_{r,\theta})$ and the PGFL of $\Xi_{l}$ \cite[Theorem 4.9]{StochasticGeometry}.
The Laplace transform (\ref{LaplacePhibLemma}) can be obtained with \cite[eq. (8)]{StochasticGeometryRandomGraphs}.
The Laplace transform (\ref{LcpsilrLemma}) can be obtained with the PGFL of the point process $\psi_b(l_{r,\theta})$
\end{IEEEproof}
The following lemma gives asymptotic analytical results of Laplace transforms in Lemma 1 under the assumption that $\lambda_l\rightarrow+\infty$, $u_b\rightarrow 0$, and the product $\lambda_l u_b$ remains constant. 
\begin{lm}
	With $\lambda_l\rightarrow+\infty$, $u_b\rightarrow 0$, and  $\lambda_l u_b=\bar{\lambda}$,  
	we have the following asymptotic  results 
	\begin{align}
	\lim_{{\lambda_l\rightarrow +\infty,u_b\rightarrow 0}}\mathcal{L}_{\Psi_b}(s)&=\mathrm{e}^{
	-2\pi\bar{\lambda}\int^{+\infty}_0{\left(1\!-\!\mathcal{L}_{\zeta_x}(sr^{-\frac{\alpha}{2}})\right)rdr}},
	\label{LaplacePSibLemmaAsymptotic}\\
     \lim_{u_b\rightarrow 0}\mathcal{L}_{\psi(l_{r})}(s)&=1.
	\label{LcpsilrLemmaAsymptotic}
	\end{align}
\end{lm}
\begin{IEEEproof}
Denoting the inner integral term in (\ref{LaplacePSibLemma}) as  $Q(r_b)=\int^{+\infty}_0\left(
1-\mathcal{L}_{\zeta_x}\left(s\left(r_b^2+t_b^2\right)^{-\frac{\alpha}{2}}\right)
\right)dt_b>0$ and employing the inequality $1+x<e^x$, we have
	\begin{align}
0<	2\lambda_l\left(1-\mathrm{exp}\left(-2\frac{\bar{\lambda}}{\lambda_l}Q(r_b)\right)\right)<{2\bar{\lambda}}Q(r_b).
	\end{align} 
It can be proved that $\int^{+\infty}_0{2\bar{\lambda}}Q(r_b)dr_b=\mathbb{E}\left(\left(s\zeta_x\right)^{\frac{2}{\alpha}}\right)\Gamma\left(1-\frac{2}{\alpha}\right)<+\infty$. Therefore, we can claim that $2\lambda_l\left(1-\mathrm{exp}\left(-2\frac{\bar{\lambda}}{\lambda_l}Q(r_b)\right)\right)$ is dominated by the integrable function ${2\bar{\lambda}}Q(r_b)$. Using the  Lebesgue's dominated convergence theorem \cite{loya2017amazing}, we can obtain (\ref{AsymptoticLaplaceTransform1}) at the top of the next page.  
	\begin{figure*}
		\begin{align}
		\lim_{\lambda_l\rightarrow +\infty,u_b\rightarrow 0}\mathcal{L}_{\Psi_b}(s)
		{=}\mathrm{exp}\left(-\int^{+\infty}_0\underset{\varpi}{\underbrace{\lim_{\lambda_l\rightarrow +\infty}2\lambda_l\left(1-\mathrm{exp}\left(-2\frac{\bar{\lambda}}{\lambda_l}Q(r_b)\right)\right)}dr_b}\right)
		\label{AsymptoticLaplaceTransform1}
		\end{align}
		\hrulefill
	\end{figure*}
	Then, employing the L'Hospital's rule, $\varpi$ in (\ref{AsymptoticLaplaceTransform1}) can be  derived as 
$\varpi=
	4\bar{\lambda} Q(r_b)$. Then, the asymptotic analytical result of $\mathcal{L}_{\Psi_b}(s)$ can be further derived as
	\begin{align}
	&\lim_{\lambda_l\rightarrow +\infty,u_b\rightarrow 0}\mathcal{L}_{\Psi_b}(s)=\mathrm{exp}\left(
	-4\bar{\lambda}\int^{+\infty}_0{Q(r_b)dr_b}\right)
	\nonumber\\
	&\overset{(a)}{=}\mathrm{exp}\left(
	-4\bar{\lambda}\int^{\frac{\pi}{2}}_0d\theta\int^{+\infty}_0{\left(1-\mathcal{L}_{\zeta_x}(sr^{-\frac{\alpha}{2}})\right)rdr}\right)
	\nonumber\\
	&=\mathrm{exp}\left(
	-2\pi\bar{\lambda}\int^{+\infty}_0{\left(1-\mathcal{L}_{\zeta_x}(sr^{-\frac{\alpha}{2}})\right)rdr}\right),
	\end{align}
where step $(a)$ is obtained due to the  polar coordinate transformation and the fact that the  integral interval is $[0,+\infty]\times[0,+\infty]$.

	Applying $u_b\rightarrow 0$ on the expression of $\mathcal{L}_{\psi(l_{r})}(s)$ given in (\ref{LcpsilrLemma}), we have $\lim_{u_b\rightarrow 0}\mathcal{L}_{\psi(l_{r})}(s)=1.$
\end{IEEEproof}
\begin{propRemark}
	From Lemma 2, the asymptotic result of the Laplace transform of the total interference power originating from vehicular networks  is 
	\begin{align}
	\mathrm{exp}\left(
	-2\bar{\lambda}\pi\int^{+\infty}_0{\left(1-\mathcal{L}_{\zeta_x}(sr^{-\frac{\alpha}{2}})\right)rdr}\right)
	\label{AsymptoticSumInterference}
	\end{align}
	Comparing (\ref{AsymptoticSumInterference}) with \cite[eq. (8)]{StochasticGeometryRadomGraph}, we can conclude that the Laplace transform of the interference power originating from vehicular networks converges to the one of the network modeled by a 2D PPP  with intensity $\bar{\lambda}$. 
\end{propRemark}

With the Slivnyak's theory \cite{StochasticGeometry}, we put a node at the origin, called the typical node to evaluate the  coverage probability and secrecy probability.  
In such case, the typical node can be, either a vehicular node or a planar node.
For the vehicular nodes, the road and the vehicular node are coupled, since every vehicular node should be on a road. The typical road is the one where the typical vehicular node locates at.
When selecting a typical vehicular node, with Palm probability, the coverage and secrecy probability analysis should be performed over \textbf{the conditional point process distribution given that a road exists at a specific location}.
By contrast, when selecting a typical planar node, we only need to consider an \textbf{unconditional} point process.  Therefore, the coverage probability and secrecy probability of a typical vehicular node  are different from the ones of a typical planar node, which will be analyzed in more detail in the forthcoming sections.

\section{Coverage Probability Analyses}
In this section, we analyze the coverage probability of the C-V2X network.
Without loss of generality, we shift the coordinate system to put the typical  user at the origin.
Since each transmitter adopts MRT, the SIR received by the typical user from the transmitter located at $x$ is
\begin{align}
\mathrm{SIR}_x = \frac{\phi||\mathbf{f}_x||_F^2D_{xo}^{-\alpha}}{\underset{{y\in\Sigma_{b/x}}}{\sum}{P_{y}}
D^{-\alpha}_{yo}
},\label{SIRxcoverage}
\end{align}
where $\Sigma_b$ is the set of interfering transmitters, and $D_{yo}$ is the distance from the transmitter at $y$ to the typical user. In (\ref{SIRxcoverage}),
 $P_{y}\triangleq\mathbf{f}_y^H\left(\frac{\phi\mathbf{f}_x\mathbf{f}_x^H}{||\mathbf{f}_x||^2_F}+\frac{(1-\phi)\mathbf{W}_x\mathbf{W}_x^H}{N-1}\right)\mathbf{f}_y$ denotes the interference power received at the
typical user from the interfering transmitter at $y\in\mathbb{R}^2$.
Since $\mathbf{f}_x\sim\mathcal{CN}\left(\mathbf{0},\mathbf{I}_N\right)$,
$||\mathbf{f}_x||_F^2\sim\mathrm{Gamma}\left(N,1\right)$, and the pdf of $P_y$ has been given in \cite[Lemma 1]{EnhancingSecrecyMultiAntenna}.

The coverage probability under the max-SIR connectivity model can be expressed as
\begin{align}
p_{c}=\mathrm{Pr}\left(\max_{x\in\Sigma_b}\mathrm{SIR}_x\geq\gamma\right).\label{pc}
\end{align}
Before commencing the coverage probability analysis, let us give the following corollary referring to the Laplace transforms of the interference power  from three types of transmitters.
\begin{corollary}
The Laplace transforms of the interference power received by the typical user at the origin, originating from the set of vehicular transmitters $\Psi_b$, from the set of planar transmitters $\Phi_b$, and from the set of vehicular transmitters $\psi_b(l_r)$  on the road with the perpendicular distance $r$, are,
respectively, given as
\begin{align}
&\mathcal{L}^c_{\Psi_b}(s)
=\mathrm{e}^{-2\lambda_l\int^{+\infty}_0\left(1-\Upsilon_u(r_b)\right)dr_b},
\label{Lcpsib}\\
&\mathcal{L}^c_{\Phi_b}(s)
=\mathrm{exp}\left(-\lambda_b\pi\omega(\phi)\Gamma(1-\delta)s^{\delta}\right),
\label{Lcphib}\\
&\mathcal{L}^c_{\psi(l_{r})}(s)
=\mathrm{e}^{-2u_b\int^{+\infty}_0\left(1-\frac{1}{\Pi\left(r^2+t_b^2\right)}\right)dt_b},
\label{Lcpsilr}
\end{align}
where
\begin{align}
&\Upsilon_u(r_b)\triangleq \mathrm{exp}\left(-2u_b\int^{+\infty}_0\left(1-\frac{1}{\Pi\left(r_b^2+t_b^2\right)}\right)dt_b\right),
\\
&\Pi\left(x\right)\!\!=\!\!{\left(\left(1\!\!+\!\!\frac{(1-\phi)s\left(x\right)^{-\frac{\alpha}{2}}}{N-1}\right)^{N-1}\right)\left(1\!\!+\!\!\phi s\left(x\right)^{-\frac{\alpha}{2}}\right)}
,
\\
&\omega(\phi)=\left\{
\begin{array}{ll}
\omega_1(\phi),&\mathrm{if}\quad \phi=\frac{1}{N},
\\
\omega_2(\phi),
&\textrm{Otherwise}.
\end{array}
\right.\label{SectorAntennaModel}
\end{align}
where $\omega_1(\phi)= \frac{\phi^{\frac{2}{\alpha}}\Gamma\left(N+\frac{2}{\alpha}\right)}{\Gamma\left(N\right)},$ and
\begin{align}
&\omega_2(\phi)=\frac{1}{\phi}\left(\frac{N-\phi^{-1}}{N-1}\right)^{1-N}
\left(\phi^{1+\frac{2}{\alpha}}\Gamma\left(1+\frac{2}{\alpha}\right)-\right.
\nonumber\\
&\left.\left(\frac{1-\phi}{N-1}\right)^{1+\frac{2}{\alpha}}\sum^{N-2}_{k=0}\left(\frac{N-\phi^{-1}}{N-1}\right)^k\frac{\Gamma\left(k+1+\frac{2}{\alpha}\right)}{\Gamma\left(k+1\right)}
\right).
\end{align}
\end{corollary}
\begin{IEEEproof}
Since $\frac{\mathbf{f}_y^H\mathbf{f}_x\mathbf{f}_x^H\mathbf{f}_y}{||\mathbf{f}_x||^2_F}\sim\mathrm{exp}(1)$ and $\frac{\mathbf{f}_y^H\mathbf{W}_x\mathbf{W}_x^H\mathbf{f}_y}{N-1}\sim\mathrm{gamma}\left(N-1,1\right)$ \cite{MultivariateStatistical}, using the Laplace transforms of  exponential and gamma random variables, the proof of (\ref{Lcpsib}) and (\ref{Lcpsilr}) can be achieved directly by employing the equations (\ref{LaplacePSibLemma}) and  (\ref{LcpsilrLemma}) in Lemma 1, which is omitted for brevity.

For proving (\ref{Lcphib}), with the equation (\ref{LaplacePhibLemma}) in Lemma 1, we have
$
\mathcal{L}^c_{\Phi_b}(s)
{=}\mathrm{exp}\left(-\lambda_b\pi\mathbb{E}\left(P_y^{\delta}\right)\Gamma(1-\delta)s^{\delta}\right),
$
and with \cite[Lemma 1]{EnhancingSecrecyMultiAntenna}, $\mathbb{E}\left(P_y^{\delta}\right)$ can be obtained as $\omega(\phi)$ given in (\ref{SectorAntennaModel}).
\end{IEEEproof}

In the following subsections, we  consider the typical user as a planar node and vehicular node, respectively, to derive the analytical result of the coverage probability in (\ref{pc}).
\subsection{Coverage Probability of the Typical Planar Receiver}
In this subsection, we consider the scenario where a planar node seating at the origin, plays as the typical user.
Further, the types of serving transmitters being independent with each other can be classified into: a) the set of planar transmitters denoted as $\Phi_b$, and b) the set of vehicular transmitters denoted as $\Psi_b$.
In this case, the coverage probability of the typical planar receiver is given in the following theorem.
\begin{Th}
Under the max-SIR connectivity model and $\gamma>1$, the coverage probability of the typical planar receiver, $\hat{p}_{c,p}$ is given by (\ref{PCPTheorem1}) at the top of the next page.
\begin{figure*}[!t]
\begin{align}
\hat{p}_{c,p}&=\mathbb{E}_{\Phi_b,\Psi_b}\left(\sum_{x\in\Phi_b+\Psi_b}\mathrm{Pr}\left(\mathrm{SIR}_x\geq \gamma\right)\right)
\!\!=\!\!2\pi\lambda_b\int^{+\infty}_0rdr
\sum^{N-1}_{n=0}\left[\frac{\left(-s_p\right)^n}{n!}\frac{d^n}{ds_p^n}\left(\mathcal{L}^c_{\Psi_b}(s_p)\mathcal{L}^c_{\Phi_b}(s_p)\right)\right]_{s_p=\gamma r^{\alpha} \phi^{-1}}
\nonumber\\
&+4\lambda_lu_b\int^{+\infty}_0dr_{b}\int^{+\infty}dt_b
\left[\sum^{N-1}_{k=0}\frac{\left(-s_v\right)^p}{p!}\frac{d^p}{ds_v^p}\left(\mathcal{L}^c_{\Phi_b}(s_v)\mathcal{L}^c_{\psi(l_{r=r_b})}(s_v) (s_v)\mathcal{L}^c_{\Psi_b}(s_v)\right)\right]_{s_v={\phi^{-1}\left(r_b^2+t_b^2\right)^{\frac{\alpha}{2}}\gamma}}.\label{PCPTheorem1}
\end{align}
\hrulefill
\end{figure*}
\end{Th}
\begin{IEEEproof}
The proof is given in Appendix A.
\end{IEEEproof}

Although the analytical result given in Theorem 1 is
exact, it is not easy for the numerical calculation, which motivates us to derive  a more easy-to-use expression  in the following theorem.
\begin{Th}
Under the max-SIR connectivity model and $\gamma>1$,  the coverage probability of the typical planar receiver, $\hat{p}_{c,p}$ can be approximated as $p_{c,p}$ in (\ref{AproximationCoverageProbabilityPlanar}) at the top of the next page,
\begin{figure*}[!t]
\begin{align}
&\hat{p}_{c,p}\lessapprox p_{c,p}= 2\pi\lambda_b\sum^{N}_{n=1}\binom{N}{n}(-1)^{n+1}\int^{+\infty}_0rdr
\left.\mathcal{L}^c_{\Psi_b}(n\kappa s_p)\mathcal{L}^c_{\Phi_b}(n\kappa s_p)\right|_{s_p=\gamma r^{\alpha} \phi^{-1}}
\nonumber\\
&+4\lambda_lu_b\sum^{N}_{n=1}\binom{N}{n}(-1)^{n+1}\int^{+\infty}_0dr_{b}\int^{+\infty}dt_b
\left.\mathcal{L}^c_{\Phi_b}\left(\kappa ns_v\right)\mathcal{L}^c_{\psi(l_{r=r_b})}(\kappa ns_v)\mathcal{L}^c_{\Psi_b}\left(\kappa ns_v\right)\right|_{s_v={\phi^{-1}\left(r_b^2+t_b^2\right)^{\frac{\alpha}{2}}\gamma}},
\label{AproximationCoverageProbabilityPlanar}
\end{align}
\hrulefill
\end{figure*}
 where $\kappa=\left(N!\right)^{-\frac{1}{N}}$.
\end{Th}
\begin{IEEEproof}
The result in (\ref{AproximationCoverageProbabilityPlanar}) can be obtained  directly by using the tight lower bound on the cdf of the gamma random variable $x$ with the scale parameter 1 and shape parameter $N$, i..e., $\mathrm{Pr}\left(x\leq y\right)\gtrapprox\left(1-e^{-\kappa y}\right)^N$ \cite{GammaInequality}.
\end{IEEEproof}

\subsection{Coverage Probability of the Typical Vehicular Receiver}
In this subsection,  we consider the scenario where a vehicular node seating at the origin, plays as the typical user, and the road passes through  the origin with $r=0$ and $\theta=0$, since the PLP is rotation-invariant.
Further, the types of serving transmitters being independent with each other can be classified into: a) the planar transmitters denoted as $\Phi_b$; b) the vehicular transmitters denoted as $\Psi_b$, and c) the vehicular transmitters on the road $l_o$ denoted as $\psi(l_o)$.
The coverage probability of the typical planar receiver is given in the following theorem. \textbf{Compared with the typical planar receiver case, the additional road $l_o$ increases the set of serving transmitters by $\psi_b(l_o)$, due to the Palm distributions}.

\begin{Th}
Under the max-SIR connectivity model and $\gamma>1$, the coverage probability of the typical vehicular receiver, $\hat{p}_{c,v}$ can be approximated by ${p}_{c,v}$ in (\ref{AproximationCoverageProbabilityVehicle}) at the top of the next page.
\begin{figure*}[!t]
\begin{align}
&\hat{p}_{c,v}\!\lessapprox\! {p}_{c,v}=
2\pi\lambda_b\sum^{N}_{n=1}\binom{N}{n}(-1)^{n+1}\int^{+\infty}_0rdr
\left.\mathcal{L}^c_{\Psi_b}(n\kappa s_p)\mathcal{L}^c_{\Phi_b}(n\kappa s_p)\mathcal{L}^c_{\psi\left(l_{r=0}\right)}\left(\kappa ns_p\right)\right|_{s_p=\gamma r^{\alpha} \phi^{-1}}
\nonumber\\
&+4\lambda_lu_b\sum^{N}_{n=1}\binom{N}{n}(\!-\!1)^{n+1}\int^{+\infty}_0dr_{b}\int^{+\infty}dt_b
\left.\mathcal{L}^c_{\Phi_b}\left(\kappa ns_v\right)\mathcal{L}^c_{\Psi_b}\left(\kappa ns_v\right)
\mathcal{L}^c_{\psi\left(l_{r\!=\!r_b}\right)}\left(\kappa ns_v\right)
\mathcal{L}^c_{\psi\left(l_{r\!=\!0}\right)}\left(\kappa ns_v\right)
\right|_{s_v={\phi^{-1}\left(r_b^2+t_b^2\right)^{\frac{\alpha}{2}}\gamma}}
\nonumber\\
&+2u_b\sum^{N}_{n=1}\binom{N}{n}(-1)^{n+1}\int^{+\infty}_0dt_b
\left.\mathcal{L}^c_{\Psi_b}(n\kappa s_{l_o})\mathcal{L}^c_{\Phi_b}(n\kappa s_{l_o})\mathcal{L}^c_{\psi\left(l_{r=0}\right)}\left(\kappa ns_{l_o}\right)\right|_{s_{l_o}={\phi^{-1}t_b^{\alpha}\gamma}}.
\label{AproximationCoverageProbabilityVehicle}
\end{align}
\hrulefill
\end{figure*}
\end{Th}
\begin{IEEEproof}
The proof is given in Appendix B.
\end{IEEEproof}

Finally, with Theorem 2 and Theorem 3, the coverage probability of the typical user is obtained by employing the total probability law and Palm probability \cite{StochasticGeometry}, which is given as follows.
\begin{corollary}
With $\gamma>1$, the coverage probability of the typical user can be approximated as
\begin{align}
p_c \thickapprox \kappa_pp_{c,p}+ \kappa_vp_{c,v},\label{pctotal}
\end{align}
where $\kappa_p\triangleq\frac{\lambda_u}{\lambda_u+u_u\lambda_l}$ denotes the probability that the typical user is a planar node, and
$\kappa_v\triangleq\frac{u_u\lambda_l}{\lambda_u+u_u\lambda_l}$ denotes the probability that the typical user is a vehicular node.
\end{corollary}

\section{Secrecy Probability Analyses}
In this section, we investigate the secrecy  probability of a typical transmitter and receiver pair.
We consider the non-colluding Eve case, and the secrecy  probability is defined as the probability that the maximal SIR received by Eves is below $\beta$, which is the SIR threshold for the secrecy outage.

We denote the channel vector between the transmitter located at $x$ and the Eve located at $e$ as $\mathbf{h}_{xe}\in\mathbb{C}^{N\times 1}$, which is a complex Gaussian random vector obeying $\mathcal{CN}\left(\mathbf{0},\mathbf{I}_N\right)$.
Since the $N\times N$ matrix $\mathbf{U}_x\triangleq\left[\frac{\mathbf{f}_x}{||\mathbf{f}_x||_F},\mathbf{W}_x\right]$ is unitary, 
the $1\times N$ vector $\mathbf{h}_{xe}^H\mathbf{U}_x=\left[q_e,\mathbf{g}^H_{xe}\right]$ where the scalar $q_e\triangleq \frac{\mathbf{h}^H_{xe}\mathbf{f}_x}{||\mathbf{f}_x||_F}$ and
the $1\times (N-1)$ vector $\mathbf{g}^H_{xe}\triangleq \mathbf{h}^H_{xe}\mathbf{W}_x$,
have independent identically distributed  (i.i.d.) complex Gaussian entries each with variance 1.
Therefore, we have $|q_e|^2\sim\mathrm{exp}(1)$ and  $||\mathbf{g}_{xe}||^2_F\sim\mathrm{gamma}\left(N-1,1\right)$
\cite{MultivariateStatistical}.
As done in \cite{EnhancingSecrecyMultiAntenna,PhysicalLayerSecurityMillimeter}, we consider the worst case by overestimating the multiuser decodability of  the spatially-distributed Eves. In particular, each Eve can estimate the CSI of the cascaded channel $\frac{\mathbf{h}_{xe}\mathbf{f}_x}{||\mathbf{f}_x||_F}$   perfectly from the pilots transmitted from each legitimate transmitter, and
 adopt the successive interference cancellation (SIC) technique
to subtract other users' information signals from the aggregate received signal, for reducing the received interference power and improving its wiretapping capability.   Hence,  when Eves try to wiretap the confidential information from the typical user, Eves only suffer the interference from the AN under the worst-case assumption.

Without loss of generality, we shift the coordinate system to put the typical  transmitter at the origin. 
Since the symbol duration is much smaller than the  coherence time of the channel,
similar to \cite{EnhancingSecrecyMultiAntenna,PhysicalLayerSecurityMillimeter},
the SIR received by the Eve at $e$ is given by
$
\mathrm{SIR}_{e}\triangleq\frac{\phi |q_e|^2D_{oe}^{-\alpha}}{\frac{(1-\phi)}{ N-1}||\mathbf{g}_{oe}||_F^2D_{oe}^{-\alpha}+{I}_e},
$
where, $\frac{(1-\phi)}{ N-1}||\mathbf{g}_{oe}||_F^2D_{oe}^{-\alpha}$ denotes the interference power originating from the typical transmitter, and ${I}_e=\underset{{{x\in\Sigma_b}}}{\sum}
\frac{(1-\phi)}{ N-1}||\mathbf{g}_{xe}||_F^2D_{xe}^{-\alpha}$ is the interference power received at the Eve from the set of transmitters, $\Sigma_b$.
The  mathematical definition of the secrecy probability $p_{sec}$ is given as
\begin{align}
p_{sec}\triangleq\mathrm{Pr}\left(\max_{e\in\Sigma_e} \mathrm{SIR}_{e}\leq\beta\right),
\end{align}
where $\Sigma_e$ denotes the set of Eves.
Before proceeding, the Laplace transforms of the interference from three types of transmitters are given in the following corollary.
\begin{corollary}
The Laplace transforms of the interference power received at the Eve from multiple vehicular transmitters, from planar transmitters and  from vehicular transmitters on the road with the perpendicular distance $r$ from the Eve, are given, respectively,  as follows
\begin{align}
&\mathcal{L}^s_{\Phi_b}(s)
{=}\mathrm{e}^{-\lambda_b\pi{\Gamma(N-1+\delta)\Gamma(1-\delta)}{\left(\Gamma(N-1)\right)^{-1}}s^{\delta}},
\\
&\mathcal{L}^s_{\Psi_b}(s)
{=}
\mathrm{e}^{-2\lambda_l\int^{+\infty}_0\left(1\!-\!\Upsilon_e(r_b)\right)dr_b},
\\
&\mathcal{L}^s_{\psi(l_{r})}(s)
=\mathrm{e}^{-2u_b\int^{+\infty}_0\left(1-{\left(1+s\left(r^2+t_b^2\right)^{-\frac{\alpha}{2}}\right)^{1-N}}\right)dt_b}.
\end{align}
where
\begin{align}
\Upsilon_e(r_b)\triangleq \mathrm{exp}\left(\!-\!2u_b\int^{+\infty}_0\left(1\!-\!{\left(1\!+\!\frac{s}{\left(r_b^2\!+\!t_b^2\right)^{\frac{\alpha}{2}}}\right)^{1\!-\!N}}\right)dt_b\right)\nonumber
\end{align}
\end{corollary}
\begin{IEEEproof}
Since $||\mathbf{g}_{xe}||^2_F\sim\mathrm{gamma}\left(N-1,1\right)$, the proof can be achieved  by employing Lemma 1 and using the Laplace transform of the gamma random variable, which is omitted.
\end{IEEEproof}

\subsection{The Typical Planar Transmitter Case}
In this subsection, we consider the scenario where a planar node seating at the origin, plays as the typical transmitter.
Further, the types of Eves being independent with each other can be classified into: a) the planar Eves denoted as $\Phi_e$, and b) the vehicular Eves denoted as $\Psi_e$.
A lower bound on the secrecy  probability is given as follows.
\begin{Th}
A lower bound on the secrecy  probability achieved by the typical planar transmitter is
\begin{align}
p^L_{sec,p}=&
\mathrm{exp}\left(-2{\lambda_l}\int^{+\infty}_0dr_e\mathrm{e}^{-2u_e\int^{+\infty}_{0}\Lambda_1\left(r_e,t_e\right)dt_e}\right)\times
\nonumber\\
&\mathrm{exp}\left(-2\pi\lambda_e \int^{+\infty}_0\Lambda_2(r_e)r_edr_e\right),
\label{pusecv}
\end{align}
where $s\triangleq\frac{(\phi^{-1}-1)\beta}{N-1}$,
$\Lambda_2\left(r_e\right)\triangleq
\left(1+s\right)^{1-N}
\mathcal{L}^s_{\Phi_b}\left(sr_e^{\alpha}\right)
\mathcal{L}^s_{\Psi_b}\left(sr_e^{\alpha})\right)$, and
$
\Lambda_1\left(r_e,t_e\right)\triangleq
\frac{\mathcal{L}^s_{\Phi_b}\left(s(r_e^2+t_e^2)^{\frac{\alpha}{2}}\right)
\mathcal{L}^s_{\Psi_b}\left(s(r_e^2+t_e^2)^{\frac{\alpha}{2}}\right)
\mathcal{L}^s_{\psi_b\left(l_{r=0}\right)}\left(s(r_e^2+t_e^2)^{\frac{\alpha}{2}}\right)}{\left(1+s\right)^{N-1}}.
$
\end{Th}
\begin{IEEEproof}
The proof is given in Appendix C.
\end{IEEEproof}

In the following,  we consider the nearest Eve only to get an upper bound on the secrecy  probability.
Before this, the pdf of the minimum distance from multiple Eves to the typical planar transmitter, is given in the following lemma.
\begin{lm}
When the nearest Eve is a vehicular node  or a planar node, the pdf of the minimum distance $d^*_e$ from multiple Eves to the typical planar transmitter
is, given by
 \begin{align}
&f_{d^*_e,\varepsilon_{0,p}}(\tau)=4\lambda_lu_e\mathrm{exp}\left(-2\lambda_l\int^{\tau}_0\left(1-e^{-2u_e\sqrt{\tau^2-r_e^2}}\right)\mathrm{d}r_e\right)
\nonumber\\
&\times\int^{\tau}_0\frac{\tau\mathrm{exp}\left(-2u_e\sqrt{\tau^2-r_e^2}\right)}{\sqrt{\tau^2-r_e^2}}\mathrm{d}r_e
\mathrm{e}^{-\lambda_e\pi \tau^2},
\label{fdevarepsilon0p}\\
&f_{d^*_e,\varepsilon_{1,p}}(\tau){=}
\mathrm{exp}\left(-2\lambda_l\int^{\tau}_0\left(1-e^{-2u_e\sqrt{\tau^2-r_e^2}}\right)\mathrm{d}r_e\right)\times
\nonumber\\
&2\lambda_e\pi \tau\mathrm{exp}\left(-\lambda_e\pi \tau^2\right).
\label{fdevarepsilon1p}
\end{align}
In the above, $\varepsilon_{0,p}$ and $\varepsilon_{1,p}$ denote the events that the nearest  Eve is a vehicular node and planar node, respectively, when the typical transmitter is a planar node.
 \end{lm}
 \begin{IEEEproof}
On the condition that the typical transmitter is a planar node, the distribution of $d^*_e$ when the nearest Eve is a vehicular node, is given by
\begin{align}
\mathrm{Pr}\left(d^*_e\geq \tau,\varepsilon_{0,p}\right)
\!\!=\!\!\mathbb{E}_{\Phi_e}\left(\int^{+\infty}_{\tau}\prod_{e\in\Phi_e}\mathbbm{l}_{||e||_2>r}
f_{d^*_e|e\in\Psi_e}(r)\mathrm{d}r\right),\label{Planarminimumde1}
\end{align}
where $f_{d^*_e|e\in\Psi_e}(r)$ is the conditional pdf of the minimum distance.
Using R$\acute{\mathrm{e}}$nyi's theorem \cite[Theorem 2.24]{StochasticGeometry} and the Leibniz¡¯ Rule, $f_{d^*_e|e\in\Psi_e}(r)$ can be derived as
\begin{align}
&f_{d^*_e|e\in\Psi_e}(r)=4\lambda_lu_e\mathrm{exp}\left(-2\lambda_l\int^{r}_0\left(1-e^{-2u_e\sqrt{r^2-r_e^2}}\right)\mathrm{d}r_e\right)
\nonumber\\
&\times\int^{r}_0\frac{r\mathrm{exp}\left(-2u_e\sqrt{r^2-r_e^2}\right)}{\sqrt{r^2-r_e^2}}\mathrm{d}r_e
.\label{Planarfde1}
\end{align}
Then, substituting (\ref{Planarfde1}) into (\ref{Planarminimumde1}), we have
\begin{align}
&f_{d^*_e,\varepsilon_{0,p}}(\tau)=-\frac{\mathrm{d} \mathrm{Pr}\left(d^*_e\geq \tau, \varepsilon_{0,p}\right)}{\mathrm{d} \tau}
{=}
-\mathbb{E}_{\Phi_e}\left(\prod_{e\in\Phi_e}\mathbbm{l}_{||e||_2>\tau}\right)
\nonumber\\
&\times
f_{d^*_e|e\in\Psi_e}(\tau)
\overset{(a)}{=}
\mathrm{exp}\left(-\lambda_e\pi \tau^2\right)f_{d^*_e|e\in\Psi_e}(\tau),
\end{align}
where step $(a)$ is due to  the PGFL of the  point processes $\Phi_e$.

The derivation of (\ref{fdevarepsilon1p})  can be achieved with a similar procedure, which is omitted for brevity.
 \end{IEEEproof}

The following lemma gives asymptotic analytical results of the minimum distance $d^*_e$ in Lemma 3 under the assumption that $\lambda_l\rightarrow+\infty$, $u_e\rightarrow 0$, and the product $\lambda_l u_e$ remains constant. 
\begin{lm}
	With $\lambda_l\rightarrow+\infty$, $u_e\rightarrow 0$, and  $\lambda_l u_e$ remains constant,  
we have the following asymptotic  results
	\begin{align}
	&\lim_{\lambda_l\rightarrow+\infty, u_e\rightarrow 0}f_{d^*_e,\varepsilon_{0,p}}(\tau)\!=\!2\lambda_lu_e\pi \tau\mathrm{exp}\left(\!-\!\left(\lambda_lu_e +\lambda_e\right) \pi \tau^2\right),\label{asymptoticMinimumDistance1}
	\\
	&\lim_{\lambda_l\rightarrow+\infty, u_e\rightarrow 0}f_{d^*_e,\varepsilon_{1,p}}(\tau)\!\!=\!\!
	2\lambda_e\pi \tau \mathrm{exp}\left(\!-\!\left(\lambda_lu_e\!\!+\!\!\lambda_e\right)\pi \tau^2\right)\label{asymptoticMinimumDistance2}
	\end{align}
\end{lm}
\begin{IEEEproof}
Following the proof of Lemma 2, the asymptotic result of $f_{d^*_e,\varepsilon_{0,p}}(d)$ can be derived as
\begin{align}
\lim_{\lambda_l\rightarrow+\infty, u_e\rightarrow 0}f_{d^*_e,\varepsilon_{0,p}}(\tau)
&\!\!=\!\!
4\lambda_l u_e\mathrm{exp}\left(\!-\!4\lambda_l u_e\int^{\tau}_0\sqrt{\tau^2\!\!-\!\!r_e^2}dr_e\right)
\nonumber\\
&\times\mathrm{e}^{-\lambda_e\pi \tau^2}\int^{\tau}_0\frac{\tau}{\sqrt{\tau^2-r_e^2}}\mathrm{d}r_e
\end{align}
Employing \cite[eq. (3.249.2) and eq. (3.248.3)]{Tableofintegrals}, we can obtain (\ref{asymptoticMinimumDistance1}). 

The derivation of (\ref{asymptoticMinimumDistance2}) can be achieved with a similar procedure which is omitted for brevity. 
\end{IEEEproof}
\begin{propRemark}
	Lemma 4 shows that the pdf of the minimum distance from multiple Eves to the typical planar transmitter converges to the one of the network modeled by a 2D PPP with intensity $\lambda_lu_e+\lambda_e$, under the assumption that  $\lambda_l\rightarrow+\infty$, $u_e\rightarrow 0$, and  $\lambda_l u_e$ remains constant.
\end{propRemark}
With Lemma 3, an upper bound on the secrecy probability is given in the following theorem.
\begin{Th}
On the condition that the typical transmitter is a planar node, by considering the nearest Eve only, an upper bound on the secrecy probability is given by
{
\begin{align}
&p_{sec,p}^U=1-
\frac{\int^{+\infty}_0\mathcal{L}^s_{\Psi_b}\left({s \tau^{\alpha}}\right)
\mathcal{L}^s_{\Phi_b}\left({s \tau^{\alpha}}\right)
f_{d^*_e,\varepsilon_{1,p}}(\tau)\mathrm{d}\tau}{\left(1+s\right)^{N-1}}-
\nonumber\\
&
\frac{\int^{+\infty}_0\mathcal{L}^s_{\Psi_b}\left({s \tau^{\alpha}}\right)
	\mathcal{L}^s_{\Phi_b}\left({s \tau^{\alpha}}\right)\mathcal{L}^s_{\psi_b(l_{r=0})}\left({s \tau^{\alpha}}\right)f_{d^*_e,\varepsilon_{0,p}}(\tau)\mathrm{d}\tau}
{\left(1+s\right)^{N-1}},
\label{psecvLTheorem}
\end{align}}
where $s\triangleq\frac{(\phi^{-1}-1)\beta}{N-1}$, $f_{d^*_e,\varepsilon_{0,p}}(\tau)$ and $f_{d^*_e,\varepsilon_{1,p}}(\tau)$ are given in (\ref{fdevarepsilon0p}) and (\ref{fdevarepsilon1p}), respectively.
\end{Th}
\begin{IEEEproof}
The proof is given in Appendix D.
\end{IEEEproof}


\subsection{The Typical Vehicular Transmitter Case}
In this subsection, we consider the scenario where a vehicular node seating at the origin, plays as the typical transmitter. The typical road $l_o$ passes through the origin with $r=0$ and $\theta=0$, since the PLP is rotation-invariant. Under the Palm distribution, the additional road $l_o$ increases the set of Eves by $\psi_e(l_o)$ , due to the Palm distributions, which would make the secrecy probability analysis  more complicated, compared with the typical planar transmitter case.
Furthermore, the types of Eves being independent with each other can be classified into: a) the planar Eves denoted as $\Phi_e$; b) the vehicular Eves on the road $l_o$ denoted as $\psi_e(l_o)$; c) the vehicular Eves on other roads denoted as $\Psi_e$.  A lower bound on the secrecy probability is given as follows.

\begin{Th}
A lower bound $p^L_{sec,v}$ on the secrecy  probability achieved by the typical vehicular transmitter, is given by
\begin{align}
p^L_{sec,v}=&
\mathrm{exp}\left(-\frac{\lambda_l}{\pi}\int^{2\pi}_0d\theta_e\int^{+\infty}_0dr_e\mathrm{e}^{-u_e\int^{+\infty}_{-\infty}\Lambda_1\left(r_e,t_e,\theta_e\right)dt_e}\right)
\nonumber\\
&\times\mathrm{exp}\left(-\lambda_e \int^{2\pi}_0\int^{+\infty}_0\Lambda_2(r_e,\theta_e)r_edr_ed\theta_e\right)
\nonumber\\
&\times\mathrm{exp}\left(-2u_e\int^{+\infty}_0\Lambda_3(t_e)dt_e\right),
\label{pusecp}
\end{align}
where $s\triangleq\frac{(\phi^{-1}-1)\beta}{N-1}$,
$h(\theta_e)\triangleq r_e\mathrm{sin}(\theta_e)-t_e\mathrm{cos}\left(\theta_e\right)$, and 
\begin{align}
&\Lambda_1\left(r_e,t_e,\theta_e\right)\triangleq
\frac{\mathcal{L}^s_{\Phi_b}\left(s(r_e^2+t_e^2)^{\frac{\alpha}{2}}\right)
\mathcal{L}^s_{\Psi_b}\left(s(r_e^2+t_e^2)^{\frac{\alpha}{2}}\right)}{\left(1+s\right)^{N-1}}
\times
\nonumber\\
&\mathcal{L}^s_{\psi_b\left(l_{r=0}\right)}\left(s(r_e^2+t_e^2)^{\frac{\alpha}{2}}\right)\mathcal{L}^s_{\psi_b\left(l_{r={h(\theta_e)}}\right)}\left(s(r_e^2+t_e^2)^{\frac{\alpha}{2}}\right),
\end{align}
\begin{align}
&\Lambda_2\left(r_e,\theta_e\right)\triangleq
\frac{\mathcal{L}^s_{\Phi_b}\left(sr_e^{\alpha}\right)
\mathcal{L}^s_{\Psi_b}\left(sr_e^{\alpha}\right)
\mathcal{L}^s_{\psi_b\left(l_{r=r_e\mathrm{sin}(\theta_e)}\right)}\left(sr_e^{\alpha}\right)}{\left(1+s\right)^{N-1}},
\nonumber\\
&\Lambda_3\left(t_e\right)\triangleq
\frac{\mathcal{L}^s_{\Phi_b}\left(st_e^{\alpha}\right)
\mathcal{L}^s_{\Psi_b}\left(st_e^{\alpha}\right)
\mathcal{L}^s_{\psi_b\left(l_{r=0}\right)}\left(st_e^{\alpha}\right)}{\left(1+s\right)^{N-1}}.
\label{Lambda3Vehicle}
\end{align}
\end{Th}
\begin{IEEEproof}
The proof is given in Appendix E.
\end{IEEEproof}

Due to  the multiple integration, the lower bound given in  Theorem 6 is computationally expensive. For alleviating the computational complexity,
we simplify the terms $\Lambda_1\left(r_e,t_e,\theta_e\right)$ and
$\Lambda_2\left(r_e,\theta_e\right)$
by removing $\mathcal{L}^s_{\psi_b\left(l_{r={h(\theta_e)}}\right)}\left(s(r_e^2+t_e^2)^{\frac{\alpha}{2}}\right)$ and $\mathcal{L}^s_{\psi_b\left(l_{r=r_e\mathrm{sin}(\theta_e)}\right)}\left(sr_e^{\alpha}\right)$, thereby obtain the following corollary.

\begin{corollary}
An easy-to-compute lower bound $\check{p}^L_{sec,v}$ on the secrecy  probability achieved by the  typical vehicular transmitter, is given by
\begin{align}
&\check{p}^L_{sec,v}=
\mathrm{exp}\left(-2{\lambda_l}\int^{+\infty}_0dr_e\mathrm{e}^{-u_e\int^{+\infty}_{-\infty}{\check{\Lambda}_1\left(r_e,t_e\right)}dt_e}\right)\times
\nonumber\\
&\mathrm{exp}\left(-2\pi\lambda_e \int^{+\infty}_0\check{\Lambda}_2(r_e)r_edr_e-2u_e\int^{+\infty}_0\Lambda_3(t_e)dt_e\right),
\label{pusecv}
\end{align}
where $s\triangleq\frac{(\phi^{-1}-1)\beta}{N-1}$, $\Lambda_3(t_e)$ is given in (\ref{Lambda3Vehicle}), and
\begin{align}
&\check{\Lambda}_1\left(r_e,t_e\right)\triangleq\left(1+s\right)^{1-N}
\mathcal{L}^s_{\Phi_b}\left(s(r_e^2+t_e^2)^{\frac{\alpha}{2}}\right)\times
\nonumber\\
&\mathcal{L}^s_{\Psi_b}\left(s(r_e^2+t_e^2)^{\frac{\alpha}{2}}\right)
\mathcal{L}^s_{\psi_b\left(l_{r=0}\right)}\left(s(r_e^2+t_e^2)^{\frac{\alpha}{2}}\right),
\nonumber\\
&\check{\Lambda}_2\left(r_e\right)\triangleq
\left(1+s\right)^{1-N}
\mathcal{L}^s_{\Phi_b}\left(sr_e^{\alpha}\right)
\mathcal{L}^s_{\Psi_b}\left(sr_e^{\alpha}\right).
\end{align}
\end{corollary}


Similar to Theorem 5,  we derive an upper bound on the secrecy probability, when considering the nearest Eve only.
As a  preliminary, the pdf of the minimum distance from multiple Eves to the typical vehicular transmitter is given in the following lemma.
\begin{lm}
 When the nearest Eve is a vehicular node but is not on the road $l_o$,
is a vehicular node on the road $l_o$, or
is a planar node, the pdf of the minimum distance $d^*_e$ from multiple Eves to the typical vehicular transmitter
is given by
 \begin{align}
&f_{d^*_e,\varepsilon_{0,v}}(\tau)=4\lambda_lu_e\mathrm{e}^{-2\lambda_l\int^{\tau}_0\left(1-e^{-2u_e\sqrt{\tau^2-r_e^2}}\right)\mathrm{d}r_e}\times
\nonumber\\
&\int^{\tau}_0\frac{\tau\mathrm{exp}\left(-2u_e\sqrt{\tau^2-r_e^2}\right)}{\sqrt{\tau^2-r_e^2}}\mathrm{d}r_e
\mathrm{e}^{-\lambda_e\pi \tau^2-2u_e \tau},
\label{fdevarepsilon0v}\\
&f_{d^*_e,\varepsilon_{1,v}}(\tau){=}
2u_e\mathrm{exp}\left(-2u_e\tau-\lambda_e\pi \tau^2\right)\times
\nonumber\\
&\mathrm{exp}\left(-2\lambda_l\int^{\tau}_0\left(1-e^{-2u_e\sqrt{\tau^2-r_e^2}}\right)\mathrm{d}r_e\right),
\label{fdevarepsilon1v}\\
&f_{d^*_e,\varepsilon_{2,v}}(\tau){=}
\mathrm{exp}\left(-2\lambda_l\int^{\tau}_0\left(1-e^{-2u_e\sqrt{\tau^2-r_e^2}}\right)\mathrm{d}r_e-2u_e \tau\right)
\nonumber\\
&\times 2\lambda_e\pi \tau\mathrm{exp}\left(-\lambda_e\pi \tau^2\right).
\label{fdevarepsilon2v}
\end{align}
Note that $\varepsilon_{0,v}$, $\varepsilon_{1,v}$, and $\varepsilon_{2,v}$, respectively, denote the event that the nearest  Eve is a vehicular node but is not on the road $l_o$, is a vehicular node on the road $l_o$, or is a planar node.
 \end{lm}
 \begin{IEEEproof}
 The proof can be achieved by following the proof of Lemma 3, which is omitted.
 \end{IEEEproof}

\begin{Th}
On the condition that the typical transmitter is a vehicular node, considering the nearest Eve only, an upper bound $p_{sec,v}^U$ on the secrecy  probability, is given by (\ref{psecvLTheorem}) at the top of the next page,
\begin{figure*}
	{
\begin{align}
p_{sec,v}^U\!=\!&1\!-\!\left(1+s\right)^{1-N}
\int^{2\pi}_0\frac{\mathrm{d}\theta_e^*}{2\pi}\int^{+\infty}_0\mathcal{L}^s_{\Psi_b}\left({s \tau^{\alpha}}\right)
\mathcal{L}^s_{\Phi_b}\left({s \tau^{\alpha}}\right)\mathcal{L}^s_{\psi_b(l_{r=0})}\left({s \tau^{\alpha}}\right)
\mathcal{L}^s_{\psi_b(l_{r=\hat{h}(\theta_e^*)})}\left({s \tau^{\alpha}}\right)f_{d^*_e,\varepsilon_{0,v}}(\tau)\mathrm{d}\tau
\nonumber\\
-&\left(1+s\right)^{1-N}
\int^{+\infty}_0\left(\mathcal{L}^s_{\Psi_b}\left({s \tau^{\alpha}}\right)
\mathcal{L}^s_{\Phi_b}\left({s \tau^{\alpha}}\right)\mathcal{L}^s_{\psi_b(l_{r=0})}\left({s \tau^{\alpha}}\right)\right)f_{d^*_e,\varepsilon_{1,v}}(\tau)\mathrm{d}\tau
\nonumber\\
-&\left(1+s\right)^{1-N}
\int^{2\pi}_0\frac{\mathrm{d}\theta_e^*}{2\pi}\int^{+\infty}_0\left(\mathcal{L}^s_{\Psi_b}\left({s \tau^{\alpha}}\right)
\mathcal{L}^s_{\Phi_b}\left({s \tau^{\alpha}}\right)
\mathcal{L}^s_{\psi_b(l_{r=\hat{h}(\theta_e^*)})}\left({s \tau^{\alpha}}\right)
\right)f_{d^*_e,\varepsilon_{2,v}}(\tau)\mathrm{d}\tau,
\label{psecvLTheorem}
\end{align}
}
\hrulefill
\end{figure*}
where $s\triangleq\frac{(\phi^{-1}-1)\beta}{N-1}$, $\hat{h}(\theta_e^*)\triangleq \tau\mathrm{sin}(\theta_e^*)$,
$f_{d^*_e,\varepsilon_{0,v}}(\tau)$, $f_{d^*_e,\varepsilon_{1,v}}(\tau)$, and $f_{d^*_e,\varepsilon_{2,v}}(\tau)$ are given in Lemma 5.
\end{Th}
\begin{IEEEproof}
The proof can be achieved by following the proof of Theorem 5. Details are omitted for brevity.
\end{IEEEproof}

It is evident that $\mathcal{L}^s_{\psi_b(l_{r={\hat{h}\left(\theta_e^*\right)}})}\left({s \tau^{\alpha}}\right)>\mathcal{L}^s_{\psi_b(l_{r={0}})}\left({s \tau^{\alpha}}\right)$,
then, an upper bound on $p_{sec,v}^U$ i.e., $\check{p}_{sec,v}^U$, can be obtained by replacing $\mathcal{L}^s_{\psi_b(l_{r={\hat{h}\left(\theta_e^*\right)}})}\left({s \tau^{\alpha}}\right)$ with $\mathcal{L}^s_{\psi_b(l_{r=0})}\left({s \tau^{\alpha}}\right)$ in (\ref{psecvLTheorem}).
Inspired by this, we build a computationally efficient upper bound, which is given in the following corollary.
\begin{corollary}
An easy to compute upper bound on the secrecy  probability achieved by the typical vehicular transmitter is given by (\ref{additionacorollary1}) at the top of the next page.
\begin{figure*}
	{
\begin{align}
\check{p}_{sec,v}^U=&1\!\!-\!\!
\frac{\int^{+\infty}_0\mathcal{L}_{\Psi_b}\left({s \tau^{\alpha}}\right)
\mathcal{L}_{\Phi_b}\left({s \tau^{\alpha}}\right)\left(\mathcal{L}_{\psi_b(l_{r=0})}\left({s \tau^{\alpha}}\right)\right)^2f_{d^*_e,\varepsilon_{0,v}}(\tau)\mathrm{d}\tau}{\left(1+s\right)^{N-1}}
\!\!-\!\!
\frac{\int^{+\infty}_0\mathcal{L}^s_{\Psi_b}\left({s \tau^{\alpha}}\right)
\mathcal{L}^s_{\Phi_b}\left({s \tau^{\alpha}}\right)\mathcal{L}^s_{\psi_b(l_{r=0})}\left({s \tau^{\alpha}}\right)f_{d^*_e,\varepsilon_{1,v}}(\tau)\mathrm{d}\tau}{\left(1+s\right)^{N-1}}
\nonumber\\
-&
\frac{\int^{+\infty}_0\mathcal{L}_{\Psi_b}\left({s \tau^{\alpha}}\right)
\mathcal{L}^s_{\Phi_b}\left({s \tau^{\alpha}}\right)
\mathcal{L}^s_{\psi_b(l_{r=0})}\left({s \tau^{\alpha}}\right)
f_{d^*_e,\varepsilon_{2,v}}(\tau)\mathrm{d}\tau}{\left(1+s\right)^{N-1}}.\label{additionacorollary1}
\end{align}
}
\hrulefill
\end{figure*}
\end{corollary}

By using total probability law \cite{StochasticGeometry}, the bounds on the secrecy probability are given as follows.
\begin{corollary}
The secrecy probability  can be bounded by
\begin{align}
\check{p}^L_{sec}\!\!\triangleq\!\!\varrho_p\check{p}^L_{sec,p}+ \varrho_v\check{p}^L_{sec,v} \leq p_c \leq \varrho_p\check{p}^U_{sec,p}+ \varrho_v\check{p}^U_{sec,v}\triangleq\check{p}^U_{sec},\label{psectotal}
\end{align}
where $\varrho_p\triangleq\frac{\lambda_b}{\lambda_b+u_b\lambda_l}$ denotes the probability that the typical transmitter is a planar node, and
$\varrho_v\triangleq\frac{u_b\lambda_l}{\lambda_b+u_b\lambda_l}$ denotes the probability that the typical transmitter is a vehicular node.
\end{corollary}



\section{Simulation Results}
\begin{figure}[!t]
\centering
\includegraphics[width=2.5in]{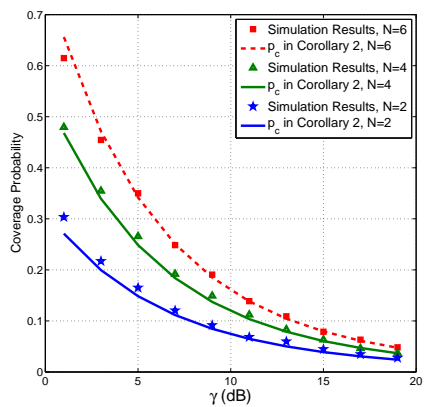}
\caption{Validation of the analytical  result of coverage probability in Corollary 2 with $\phi=0.6, \alpha=2.3,\lambda_b=10^{-5}, \lambda_l=5\times 10^{-4}, u_b=10^{-3}$.}
\label{CoverageProbabilityVehicleUser}
\end{figure}

In this section, the simulation results of the coverage probability and secrecy probability are provided for validating the  theoretical results in Corollary 2 and Corollary 6. Then, we evaluate the impact of the network parameters on the secrecy performance of the C-V2X network by simulations.
Fig. \ref{CoverageProbabilityVehicleUser} shows the approximate coverage probability given in (\ref{pctotal}) and the simulation results versus the SIR threshold $\gamma$ for different numbers of antennas, $N$. Simulation results validate the accuracy of the analytical result, and show that the approximate analytical result in (\ref{pctotal}) coincides with the simulation results very well, when $N$ increases from 2 to 6. 


\begin{figure}[!t]
	\centering
	\includegraphics[width=2.5in]{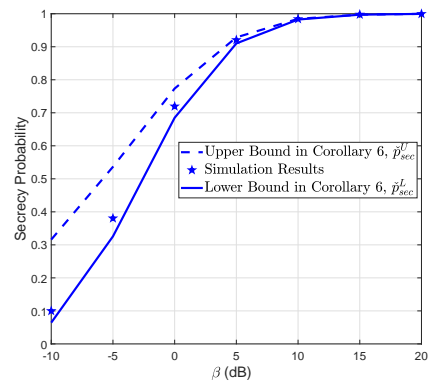}
	\caption{Validation of the tightness of the lower bound and upper bound in Corollary 8 with parameters
		$N=2,\phi=0.6,\lambda_l=10^{-4},u_b=10^{-4},u_e=10^{-4},\lambda_b=10^{-5},\lambda_e=10^{-4}$.}
	\label{SecrecyOutageProbability}
\end{figure}

\begin{figure}[!t]
	\centering
	\includegraphics[width=2.5in]{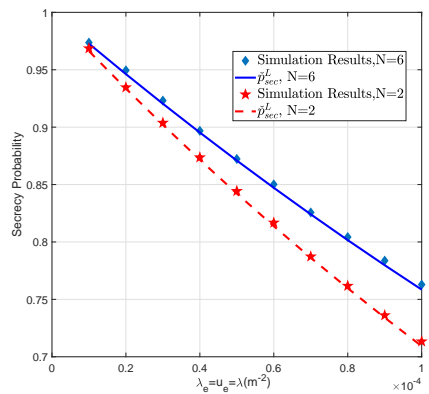}
	\caption{Validation of the lower bound $\check{p}^L_{sec}$  versus the intensity $\lambda$ for
		$\phi=0.6,\alpha=3,\lambda_l=u_b=\lambda_b=10^{-4}$, $\lambda_e=u_e=\lambda$.}
	\label{SecrecyOutageVSlambda}
\end{figure}



In Fig. \ref{SecrecyOutageProbability}, we plot the simulation results of the secrecy  probability versus the bounds given in Corollary 6.
From Fig. \ref{SecrecyOutageProbability},  we can find  that $\check{p}_{sec}^L$ coincides with the simulation results very well, which validates the tightness of the lower bound $\check{p}_{sec}^L$.
The upper bound $\check{p}_{sec}^U$ is loose, especially when the SIR threshold is lower than 0 dB.


For validating the tightness of the lower bound $\check{p}_{sec}^L$ further, we plot the simulation results of the secrecy  probability  $\check{p}_{sec}^L$ versus $\lambda$ with $\lambda_e=u_e=\lambda$ in Fig. \ref{SecrecyOutageVSlambda}.
With the increasing $\lambda$, the number of Eves eavesdropping the confidential information from the typical vehicular transmitter increases, and the secrecy  probability decreases. From the simulation results in Fig. \ref{SecrecyOutageVSlambda}, we can find that $\check{p}_{sec}^L$ coincides with the simulation results very well over the whole range of $\lambda$, which has validated the tightness of the lower bound further.

Since the simulation results in Fig. \ref{CoverageProbabilityVehicleUser} and Fig. \ref{SecrecyOutageVSlambda} have validated the analytical results given in Corollary 2 and Corollary 6, respectively, according to Definition 1, the analytical result of the effective secrecy throughput is
\begin{align}
\eta = R_s\left(\kappa_pp_{c,p}+ \kappa_vp_{c,v}\right)\check{p}^L_{sec}.\label{ApproximateSecrecyThroughput}
\end{align}
Setting $\gamma=10$dB and  $\beta=0$dB, $R_s=\mathrm{log}_2\left(1+\gamma\right)-\mathrm{log}_2\left(1+\beta\right)$ and the effective secrecy throughput is given  by (\ref{ApproximateSecrecyThroughput}). The following simulation results of the effective secrecy throughput are  theoretical results given in (\ref{ApproximateSecrecyThroughput}).

\begin{figure}[!t]
	\centering
	\includegraphics[width=2.5in]{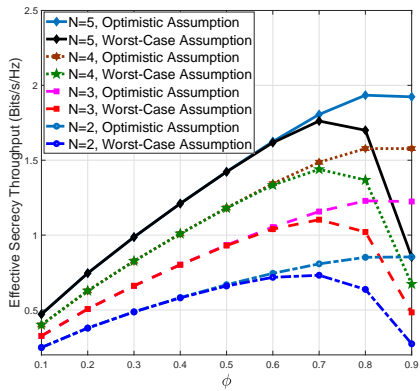}
	\caption{{The effective secrecy  throughput of the typical transmitter and receiver pair versus $\phi$ for	$N=2,\alpha=2.3, \lambda_b=10^{-5},\lambda_l=5\times10^{-4},
			u_b=u_e=10^{-3},\lambda_e=10^{-4}
			$;}}
	\label{SecrecyThroughputVSPhi}
\end{figure}

Fig. \ref{SecrecyThroughputVSPhi} plots the effective secrecy throughput versus the power allocation ratio $\phi$ for the worst-case assumption adopted in this work and an optimistic assumption. 
For the worst-case assumption, this work performs the secrecy performance analysis by overestimating  the multi-user decodability of Eves. Instead, a more optimistic assumption can be built by underestimating the multi-user decodability of Eves, where both  information signals and AN transmitted by transmitters act as the interference to deteriorate the wiretapping capability of Eves.
First, for the worst-case assumption, Fig. \ref{SecrecyThroughputVSPhi} shows that the effective secrecy throughput may be a concave/quais-concave function of $\phi$, and there is a unique optimal $\phi$ for maximizing the effective secrecy throughput. The optimal $\phi\approx 0.7$, which means that  most of the power is allocated to the confidential information. Furthermore, the simulation results show that the optimal $\phi$ remains unchanged approximately, with the increasing number of antennas.
Second, for the optimistic assumption, with the increasing $\phi$, the C-V2X network gradually achieves a better secrecy performance than the one under the worst-case assumption. Furthermore, with the increasing $\phi$, its achievable secrecy throughput converges to a steady constant, which shows that even for the optimistic assumption, increasing the information signal power may not always benefit the secrecy performance due to the information leakage. 

\begin{figure}[!t]
	\centering
	\includegraphics[width=2.5in]{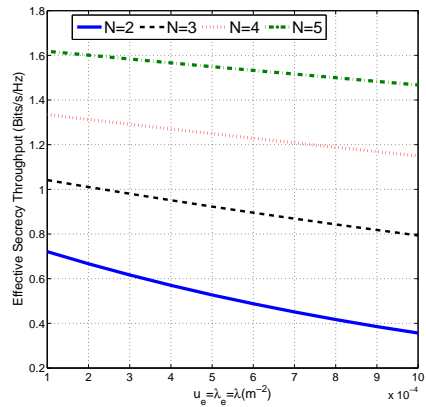}
	\caption{The effective secrecy  throughput of the typical transmitter and receiver pair versus the intensity of eavesdroppers $\lambda$ for
		$N=5,\phi=0.6,\alpha=2.3$, $\lambda_b=10^{-5},\lambda_l=5\times 10^{-4},u_b=10^{-3},\lambda_e=\lambda,u_e=5\times\lambda
		$.}
	\label{SecrecyThroughputVSlambdaTheoretical}
\end{figure}



Fig. \ref{SecrecyThroughputVSlambdaTheoretical} shows the change trend of the effective secrecy throughput with the increasing intensity of Eves.
Obviously, the effective secrecy throughput decreases with the increasing $\lambda$. When $N=2$, the effective secrecy throughput decreases by 0.37 bits/s/Hz for $\lambda$ changing from $10^{-4}$ to $10^{-3}$. But, when $N=5$, the effective secrecy throughput only decreases by about 0.15 bits/s/Hz. Then, we can make a conclusion that the  decreasing rate of the effective secrecy throughput decreases with the increasing $N$, and a larger antenna array can improve the robustness of the secure transmission scheme.

\begin{figure}[!t]
	\centering
	\includegraphics[width=2.5in]{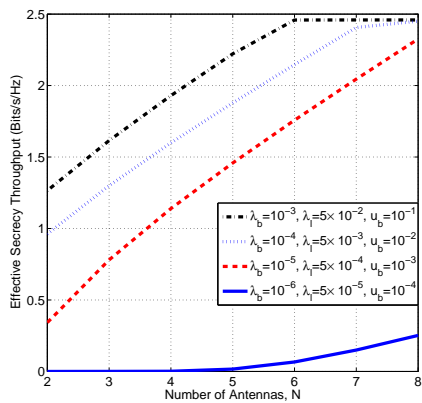}
	\caption{The effective secrecy  throughput of the typical transmitter and receiver pair versus the number antennas equipped at each transmitter $N$ for
		$\phi=0.6,\alpha=2.3$, $\lambda_e=10^{-3},u_e=10^{-2}$;}
	\label{SecrecyThroughputVSNumberofAntenna}
\end{figure}

Fig. \ref{SecrecyThroughputVSNumberofAntenna} shows  the change trend of the effective secrecy throughput with the increasing number of antennas $N$. Just as the simulation results in Fig. \ref{SecrecyThroughputVSlambdaTheoretical}, more numbers of antennas employed would result in a better secrecy performance. But, the top simulation curves in Fig. \ref{SecrecyThroughputVSNumberofAntenna} show that the  growing rate of the effective secrecy throughput decreases when $N$ increases from 6 to 8. This shows that the effective secrecy throughput can not increase linearly with $N$ all the time and there is a best tradeoff between improving the  secrecy performance and the system complexity.

\begin{figure}[!t]
	\centering
	\includegraphics[width=2.5in]{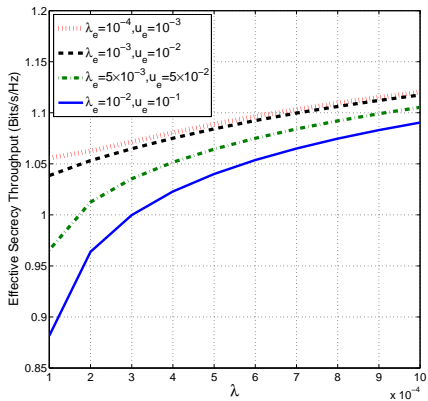}
	\caption{The effective secrecy  throughput of the typical transmitter and receiver pair versus the intensity of transmitters for $N=4,\lambda_b=\lambda$, $u_b=5\lambda$
		$\phi=0.6,\alpha=2.3$, $\lambda_l=5\times10^{-4}$.}
	\label{SecrecyThroughputVSIntensityofBaseStation}
\end{figure}



Fig. \ref{SecrecyThroughputVSIntensityofBaseStation} shows the effective secrecy throughput versus the intensity of transmitters. With the increasing intensity of transmitters, the  effective secrecy throughput  increases, due to the decreasing distance between transmitter and receiver, and the increasing power of AN.
But, from the simulation results, we can find that the secrecy performance gain obtained by increasing intensity of transmitters is small. Furthermore, when the intensity of Eves is small, the secrecy performance gains brought by making the network denser, is lower than the network with a larger intensity of Eves.


\section{Conclusion}
In this paper, we built a theoretical framework for analyzing the PLS of AN assisted C-V2X networks by leveraging the Cox point process.  Closed-form analytical expressions of  coverage probability and secrecy probability were derived, which facilitate the evaluation of the effective secrecy throughput. The simulation results show the impact of network parameters on the achievable secrecy performance. In particular, the effective secrecy throughput may be a concave/quasi-concave function of the  power allocation coefficient, and
the capacity for suppressing eavesdropping increases with the  number of transmit antennas.

This work may have  some potential extensions  as follows. First,  the spatial distribution of vehicles over a road is built by a static PPP model in this work. It is interesting to extend the spatial model to  a stochastic geometry-based mobility model for characterizing the impact of the mobility of vehicles on the secrecy performance. Moreover, in this work, we just studied the secrecy performance of  unicast communication. In practical systems, different communication modes coexist. Therefore, it is important to extend the secrecy performance analysis to a  network with multiple communication modes, e.g., broadcast, unicast and multicast communications.

\appendices
\section{Proof of Theorem 1}
On the condition that the typical user is a planar node, the coverage probability in (\ref{pc}) is given by
\begin{align}
\hat{p}_{c,p}\!\!=\!\!\underset{\hat{p}_{c,p,1}}{\underbrace{\mathbb{E}_{\Phi_b}\left(\sum_{x\in{\Phi_b}}\mathrm{Pr}\left(\mathrm{SIR}_x\!\!\geq\!\! \gamma\right)\right)}}\!\!+\!\!
\underset{\hat{p}_{c,p,2}}{\underbrace{\mathbb{E}_{\Psi_b}\left(\sum_{x\in{\Psi_b}}\mathrm{Pr}\left(\mathrm{SIR}_x\!\!\geq\!\! \gamma\right)\right)}}.\label{pcp}
\end{align}

The first part of (\ref{pcp}), $\hat{p}_{c,p,1}$, is the coverage probability when the typical user is associated with a planar transmitter, which can be derived as (\ref{ConnectionOutageA1}) at the top of the next page, where
\begin{figure*}
\begin{align}
\hat{p}_{c,p,1}&\!\!\overset{(a)}{=}\!\!2\pi\lambda_b\int^{+\infty}_0\mathrm{Pr}\left(\frac{\phi r^{-\alpha}||\mathbf{f}_x||_F^2}{\underset{{y\in\left\{\Psi_b+\Phi_b\right\}}}{\sum}{P_y}{D^{-\alpha}_{yo}}}\!\!\geq\!\! \gamma\right)rdr
\!\!\overset{(b)}{=}\!\!2\pi\lambda_b\int^{+\infty}_0
\sum^{N-1}_{n=0}\left[\frac{\left(-s_p\right)^n}{n!}\frac{d^n}{ds_p^n}\left(\mathcal{L}^c_{\Psi_b}(s_p)\mathcal{L}^c_{\Phi_b}(s_p)\right)\right]_{s_p=\gamma r^{\alpha} \phi^{-1}}rdr,
\label{ConnectionOutageA1}
\end{align}
\hrulefill
\end{figure*}
step ($a$) is obtained by employing Campbell's theorem for sums over the PPP $\Phi_b$ \cite[Theorem 4.1]{StochasticGeometry} and step ($b$) is due to \cite[Theorem 1]{AdHocSpatialDiversity}.
The second part of (\ref{pcp}), $\hat{p}_{c,p,2}$, is the coverage probability when the typical user is associated with a vehicular transmitter, which can be derived as (\ref{ConnectionOutageA2}) at the top of the next page.
\begin{figure*}
\begin{align}
\hat{p}_{c,p,2}&=\mathbb{E}_{\Xi_l}\left(\sum_{\left\{r_b,\theta_b\right\}\in\Xi_l}\mathbb{E}\left(\sum_{t_b\in l_{r_b,\theta_b}}\mathrm{Pr}\left(\frac{\phi \left(r_b^2+t_b^2\right)^{-\frac{\alpha}{2}}||\mathbf{f}_x||_F^2}{\underset{{y\in\left\{\Psi_b+\Phi_b+\psi(l_{r=r_b})\right\}}}{\sum}{P_y}D^{-\alpha}_{yo}}\geq \gamma\right)\right)\right)
\nonumber\\
&{=}4\lambda_lu_b\int^{+\infty}_0dr_b\int^{+\infty}_0dt_b
\sum^{N-1}_{n=0}\left[\frac{\left(-s_v\right)^n}{n!}\frac{d^n}{ds_v^n}\left(\mathcal{L}^c_{\Psi_b}(s_v)\mathcal{L}^c_{\Phi_b}(s_v)\mathcal{L}^c_{\psi_b(l_{r=r_b})}(s_v)\right)\right]_{s_v={\phi^{-1}\left(r_b^2+t_b^2\right)^{\frac{\alpha}{2}}\gamma}}.
\label{ConnectionOutageA2}
\end{align}
\hrulefill
\end{figure*}

\section{Proof of Theorem 3}
On the condition that the typical user is a vehicular node, the analytical result of the coverage probability in (\ref{pc}) can be derived as
\begin{align}
\hat{p}_{c,v}&\!=\!
\underset{\hat{p}_{c,v,1}}{\underbrace{\mathbb{E}_{\Phi_b}\left(\sum_{x\in{\Phi_b}}\mathrm{Pr}\left(\mathrm{SIR}_x\geq \gamma\right)\right)}}\!+\!
\underset{\hat{p}_{c,v,2}}{\underbrace{\mathbb{E}_{\Psi_b}\left(\sum_{x\in{\Psi_b}}\mathrm{Pr}\left(\mathrm{SIR}_x\geq \gamma\right)\right)}}
\nonumber\\
&\!+\!
\underset{\hat{p}_{c,v,3}}{\underbrace{\mathbb{E}_{\psi_b(l_{r=0})}\left(\sum_{x\in{\psi_b(l_{r=0})}}\mathrm{Pr}\left(\mathrm{SIR}_x\geq \gamma\right)\right)}}
.
\label{pcv}
\end{align}

The first part of (\ref{pcv}), $\hat{p}_{c,v,1}$, is the coverage probability when the typical user is associated with a planar transmitter. Following the derivation in (\ref{ConnectionOutageA1}) and defining $\Omega_1\triangleq \Psi_b\!+\!\Phi_b\!+\!\psi\left(l_{r=0}\right)$,  we have
\begin{align}
\hat{p}_{c,v,1}&\!\!=\!\!2\pi\lambda_b\int^{+\infty}_0\mathrm{Pr}\left(\frac{\phi r^{-\alpha}||\mathbf{f}_x||_F^2}{\underset{y\in\Omega_1}{\sum}{P_y}D^{-\alpha}_{yo}}\!\geq\! \gamma\right)rdr.
\label{ConnectionOutageVehicle1}
\end{align}

The second part of (\ref{pcv}), $\hat{p}_{c,v,2}$, is the coverage probability when the typical user is associated with a vehicular transmitter. Following the derivation in (\ref{ConnectionOutageA2}) and
defining $\Omega_2\triangleq \Psi_b+\Phi_b+\psi(l_{r=r_b})+\psi(l_{r=0})$, 
 we have
\begin{align}
\hat{p}_{c,v,2}\!=\!4\lambda_lu_b\int^{+\infty}_0dr_b\int^{+\infty}_0\mathrm{Pr}\left(\frac{\frac{\phi} {\left(r_b^2+t_b^2\right)^{\frac{\alpha}{2}}}||\mathbf{f}_x||_F^2}{\underset{{y\in\Omega_2}}{\sum}{P_y}{D^{-\alpha}_{yo}}}\!\geq\! \gamma\right)dt_b.
\end{align}

The third part of (\ref{pcv}), $\hat{p}_{c,v,3}$, is the coverage probability when the typical user is associated with a vehicular transmitter on the road $l_o$. Employing the Campbell's theorem for sums over the point process $\psi_b\left(l_{r=0}\right)$ \cite[Theorem 4.1]{StochasticGeometry} and defining
$\Omega_3\triangleq \Psi_b+\Phi_b+\psi\left(l_{r=0}\right)$,  we have
\begin{align}
\hat{p}_{c,v,3}&{=}2u_b\int^{+\infty}_0\mathrm{Pr}\left(\frac{\phi t_b^{-\alpha}||\mathbf{f}_x||_F^2}{\underset{{y\in\Omega_3}}{\sum}{P_y}D^{-\alpha}_{yo}}\geq \gamma\right)dt_b.
\label{ConnectionOutageVehicle1}
\end{align}

Finally, employing Corollary 1, the exact analytical result is given as (\ref{CoverageProbabilityVehicularReceiver}) at the top of the next page.
\begin{figure*}
\begin{align}
&p_{c,p}=4\lambda_lu_b\int^{+\infty}_0dr_{b}\int^{+\infty}dt_b
\left[\sum^{N-1}_{n=0}\frac{\left(-s_v\right)^n d^n}{n!ds_v^n}\left(\mathcal{L}^c_{\Phi_b}(s_v)\mathcal{L}^c_{\psi(l_{r=r_b})}(s_v) \mathcal{L}^c_{\psi(l_{r=0})} (s_v)\mathcal{L}^c_{\Psi_b}(s_v)\right)\right]_{s_v={\phi^{-1}\left(r_b^2+t_b^2\right)^{\alpha/2}\gamma}}
\nonumber\\
&+2\pi\lambda_b\int^{+\infty}_0rdr
\sum^{N-1}_{n=0}\left[\frac{\left(-s_p\right)^n}{n!}\frac{d^n}{ds_p^n}\left(\mathcal{L}^c_{\Psi_b}(s_p)\mathcal{L}^c_{\psi(l_{r=0})}(s_p) \mathcal{L}^c_{\Phi_b}(s_p)\right)\right]_{s_p={\phi^{-1}r^{\alpha}\gamma}}
\nonumber\\
&+2u_b\int^{+\infty}_0dt_b
\sum^{N-1}_{n=0}\left[\frac{\left(-s_{l_o}\right)^n}{n!}\frac{d^n}{ds_{l_o}^n}\left(\mathcal{L}^c_{\Psi_b}(s_{l_o})\mathcal{L}^c_{\psi(l_{r=0})}(s_{l_o}) \mathcal{L}^c_{\Phi_b}(s_{l_o})\right)\right]_{s_{l_o}={\phi^{-1}t_b^{\alpha}\gamma}}.\label{CoverageProbabilityVehicularReceiver}
\end{align}
\hrulefill
\end{figure*}
Then, just  as deriving the approximate result in Theorem 2, employing
the lower bound on the cdf of the gamma random variable \cite{GammaInequality}, the approximate result given in (\ref{AproximationCoverageProbabilityVehicle}) can be obtained directly.
\section{Proof of Theorem 4}
The secrecy  probability can be expressed as
\begin{spacing}{.5}
\begin{align}
&\mathrm{Pr}\left(\max_{e\in\left(\Phi_{e}+\Psi_{e}\right)}\mathrm{SIR}_e\leq \beta\right)
=\mathbb{E}\left(
\underset{(\mathrm{I})}{\underbrace{\prod_{e\in\Psi_{e}}\mathrm{Pr}\left(\mathrm{SIR}_e\leq\beta\right)}}\times\right.
\nonumber\\
&\left.\underset{(\mathrm{II})}{\underbrace{\prod_{e\in\Phi_{e}}\mathrm{Pr}\left(\mathrm{SIR}_e\leq\beta\right)}}\right).
\label{SececyOutagePlanar}
\end{align}
\end{spacing}
The expectation over the product $(\mathrm{I})$ can be derived as (\ref{productIPlanar}) at the top of the next page,
\begin{figure*}
\begin{align}
&\mathbb{E}\left(\prod_{e\in\Psi_{e}}\mathrm{Pr}\left(\mathrm{SIR}_e\leq\beta\right)\right)
\!\!\overset{(a)}{=}\!\!\mathbb{E}\left(\prod_{(r_e,\theta_e)\in\Xi_{l}}\mathbb{E}\left(\prod_{t_e\in\psi(l_{r_e,\theta_e})}\mathrm{Pr}\left(\left.\frac{\phi |q_e|^2\left(r_e^2+t_e^2\right)^{-\frac{\alpha}{2}}}{
\frac{(1-\phi)||\mathbf{g}_{oe}||_F^2\left(r_e^2+t_e^2\right)^{-\frac{\alpha}{2}}}{N-1}+\underset{{x\in\Delta_{b,p}}}{\sum}\frac{(1-u)||\mathbf{g}_{xe}||_F^2D_{xe}^{-\alpha}}{N-1}}\!\!\leq\!\!\beta\right|\Delta_{b,p}\right)\right)\right)
\nonumber\\
&\overset{(b)}{=}\mathbb{E}\left(\mathrm{exp}\left(-\frac{\lambda_l}{\pi}\int^{2\pi}_0d\theta_e\int^{+\infty}_0dr_e\mathrm{e}^{-2u_e\int^{+\infty}_{0}\mathrm{Pr}\left(\left.\frac{\frac{\phi\left(N-1\right)}{1-\phi}|q_e|^2\left(r_e^2+t_e^2\right)^{-\frac{\alpha}{2}}}{
||\mathbf{g}_{oe}||_F^2\left(r_e^2+t_e^2\right)^{-\frac{\alpha}{2}}+\underset{{x\in\Delta_{b,p}}}{\sum}||\mathbf{g}_{xe}||_F^2D_{xe}^{-\alpha}}\geq\beta\right|\Delta_{b,p}\right)dt_e}\right)\right)
\nonumber\\
&\overset{(c)}{\gtrapprox}
\mathrm{exp}\left(-{2\lambda_l}\int^{+\infty}_0dr_e\mathrm{e}^{-2u_e\int^{+\infty}_{0}\mathrm{Pr}\left(\frac{\frac{\phi\left(N-1\right)}{1-\phi}|q_e|^2\left(r_e^2+t_e^2\right)^{-\frac{\alpha}{2}}}{
||\mathbf{g}_{oe}||_F^2\left(r_e^2+t_e^2\right)^{-\frac{\alpha}{2}}+\underset{{x\in\Delta_{b,p}}}{\sum}||\mathbf{g}_{xe}||_F^2D_{xe}^{-\alpha}}\geq\beta\right)dt_e}\right),
\label{productIPlanar}
\end{align}
\hrulefill
\end{figure*}
where $\Delta_{b,p}\triangleq{\Phi_b\cup\Psi_b\cup\psi_b(l_{e})}$.
At step $(a)$, the distance between the typical planar transmitter and the vehicular Eve is $\sqrt{r_e^2+t_e^2}$, which is illustrated in Fig. \ref{NetworkTopology}, where $r_e$ is the perpendicular distance from the typical transmitter located at the origin to the road, and $t_e$ denotes the distance of the vehicular Eve from the projection of the origin onto the road, i.e., $o'$.
Notice that under the condition that $e\in\Psi_e$, the eavesdropper is on a road which is denoted as $l_e$ and the corresponding point on the representation space is denoted as $(r_e,\theta_e)$. Therefore, for each vehicular Eve at $e\in\Psi_e$, the interference received at Eve comes from multiple planar transmitters, vehicular transmitters on the road $l_e$, and vehicular transmitters on the other roads. Step $(b)$ is obtained by using the PGFL of the point process $\psi(l_{r_e,\theta_e})$ and the point process $\Xi_{l}$ \cite[Theorem 4.9]{StochasticGeometry}. Step ($c$) is due to the Jensen's inequality.

Employing the PGFL of $\Phi_{e}$,
the expectation over the product $(\mathrm{II})$ can be derived as
\begin{align}
&\mathbb{E}\left(\prod_{e\in\Phi_{e}}\mathrm{Pr}\left(\mathrm{SIR}_e\leq\beta\right)\right)
{=}
\nonumber\\
&e^{-2\lambda_e \pi\int_0^{+\infty}\mathrm{Pr}\left(\frac{\frac{\phi(N-1)}{1-\phi}|q_e|^2r_e^{-{\alpha}}}{
 ||\mathbf{g}_{oe}||_F^2r_e^{-{\alpha}}+\underset{{x\in\{\Phi_b+\Psi_b\}}}{\sum}{||\mathbf{g}_{xe}||_F^2D_{xe}^{-\alpha}}}\geq\beta\right)r_edr_e}.
\label{productIIPlanar}
\end{align}
Since the planar Eve does not have to locate at any road, the planar Eve only suffers the interference from the transmitters at $\Phi_b\cup\Psi_b$.

For obtaining the secrecy  probability, the complementary cumulative distribution function  (ccdf) of the SIR received by Eve should be derived first.
Defining $s\triangleq\frac{(\phi^{-1}-1)\beta}{N-1}$, the ccdf in (\ref{productIPlanar}) is given as (\ref{ccdfSINRePlanar}) at the top of the next page.
\begin{figure*}
\begin{align}
&\mathrm{Pr}\left(\frac{\phi |q_e|^2\left(r_e^2+t_e^2\right)^{-\frac{\alpha}{2}}}{
\frac{1-\phi}{\left(N-1\right)\left(r_e^2+t_e^2\right)^{\frac{\alpha}{2}}}||\mathbf{g}_{oe}||_F^2+\underset{{x\in\{\Phi_b+\Psi_b+\psi(l_e)\}}}{\sum}\frac{(1-\phi)||\mathbf{g}_{xe}||_F^2D_{xe}^{-\alpha}}{(N-1)}}\geq\beta\right)
\overset{(a)}{=}\frac{\mathrm{exp}\left(-\left(s\left(r_e^2+t_e^2\right)^{\frac{\alpha}{2}}\underset{{x\in\{\Phi_b+\Psi_b+\psi(l_e)\}}}{\sum}||\mathbf{g}_{xe}||_F^2D_{xe}^{-\alpha}\right)\right)}{\left(1+\beta\frac{\phi^{-1}-1}{N-1}\right)^{N-1}}
\nonumber\\
&\overset{(b)}{=}\left(1+\beta\frac{\phi^{-1}-1}{N-1}\right)^{1-N}
\underset{(\mathrm{i})}{{\mathcal{L}^s_{\Phi_b}\left(s\left(r_e^2+t_e^2\right)^{\frac{\alpha}{2}}\right)}}
\underset{(\mathrm{ii})}{\mathcal{L}^s_{\Psi_b}\left(s\left(r_e^2+t_e^2\right)^{\frac{\alpha}{2}}\right)}
\underset{(\mathrm{iii})}{\mathcal{L}^s_{\psi_b(l_{r=0})}\left(s\left(r_e^2+t_e^2\right)^{\frac{\alpha}{2}}\right)}.
\label{ccdfSINRePlanar}
\end{align}
\hrulefill
\end{figure*}
Step $(a)$ is due to the Laplace transform of the gamma random variable, and step $(b)$ is due to that $\Phi_b$, $\Psi_b$,  and $\psi(l_{r=0})$ are all independent.

Applying  Corollary 3,   Laplace transforms (i)-(iii) in (\ref{ccdfSINRePlanar}) can be derived. Substituting the analytical result of (\ref{ccdfSINRePlanar}) into (\ref{productIPlanar}), the analytical result of $\mathbb{E}\left(\underset{{e\in\Psi_{e}}}{\prod}\mathrm{Pr}\left(\mathrm{SIR}_e\leq\beta\right)\right)$ can be obtained.
With the same procedures, the analytical result of $\mathbb{E}\left(\underset{{e\in\Phi_{e}}}{\prod}\mathrm{Pr}\left(\mathrm{SIR}_e\leq\beta\right)\right)$ can be obtained. Then, the secrecy  probability can be derived with (\ref{SececyOutagePlanar}).

\section{Proof of Theorem 5}
On the condition that the typical transmitter is a planar node, the secrecy  probability is derived by considering two events: 1) the nearest  Eve is a vehicular node, which is denoted by $\varepsilon_{0,p}$; 2) the nearest Eve is a planar node, which is denoted as $\varepsilon_{1,p}$. Then, considering the nearest Eve only, the secrecy  probability $p_{sec,p}^L$ can be established as (\ref{psecvLPlanar}) at the previous page.
\begin{figure*}[!t]
\begin{align}
p_{sec,p}^L&=\underset{p_{sec,p,1}^L}{\underbrace{\mathbb{E}_{d_e^*}\left(\mathrm{Pr}\left(\frac{\phi |q_e|^2\left({d^*_e}\right)^{-\alpha}}{\frac{(1-\phi)||\mathbf{g}_{oe}||_F^2\left({d^*_e}\right)^{-\alpha}}{N-1}+\underset{{x\in\left\{\Phi_b+\Psi_b+\psi_b(l_e)\right\}}}{\sum}\frac{(1-\phi)||\mathbf{g}_{xe}||_F^2D_{xe}^{-\alpha}}{N-1}}\leq \beta,\varepsilon_{0,p}\right)\right)}}\nonumber\\
&+\underset{p_{sec,p,2}^L}{\underbrace{\mathbb{E}_{d^*_e}\left(\mathrm{Pr}\left(\frac{\phi |q_e|^2\left({d^*_e}\right)^{-\alpha}}{\frac{(1-\phi)||\mathbf{g}_{oe}||_F^2\left({d^*_e}\right)^{-\alpha}}{N-1}+\underset{{x\in\left\{\Phi_b+\Psi_b\right\}}}{\sum}\frac{(1-\phi)||\mathbf{g}_{xe}||_F^2D_{xe}^{-\alpha}}{N-1}}\leq \beta,\varepsilon_{1,p}\right)\right)}}.
\label{psecvLPlanar}
\end{align}
\hrulefill
\end{figure*}

Let's derive the first expectation $p_{sec,v,1}^L$ in (\ref{psecvLPlanar}).
\begin{align}
&p_{sec,p,1}^L=\!\!\int^{+\infty}_0\!\!\left(1-\frac{\mathcal{L}^s_{\Psi_b}\left({s \tau^{\alpha}}\right)
	\mathcal{L}^s_{\Phi_b}\left({s \tau^{\alpha}}\right)\mathcal{L}^s_{\psi_b(l_{r=0})}\left({s \tau^{\alpha}}\right)}{\left(1+s\right)^{N-1}}\right)
\nonumber\\
&\times
f_{d^*_e,\varepsilon_{0,p}}(\tau)\mathrm{d}\tau,
\label{psecvL1Planar}
\end{align}
where $f_{d^*_e,\varepsilon_{0,p}}(\tau)$ is  given in (\ref{fdevarepsilon0p}).
With Corollary 3, the analytical result of $p_{sec,v,1}^L$ can be obtained.
Second, let's derive the second expectation $p_{sec,p,2}^L$ in (\ref{psecvLPlanar}).
\begin{align}
p_{sec,p,2}^L=&\int^{+\infty}_0\left(1-
\frac{\mathcal{L}^s_{\Psi_b}\left({s \tau^{\alpha}}\right)
\mathcal{L}^s_{\Phi_b}\left({s \tau^{\alpha}}\right)}{\left(1+s\right)^{N-1}}\right)
f_{d^*_e,\varepsilon_{1,p}}(\tau)\mathrm{d}\tau,\label{psecvL2Planar}
\end{align}
where $f_{d^*_e,\varepsilon_{1,p}}(\tau)$ has been given in (\ref{fdevarepsilon1p}).
Employing Corollary 3, the analytical result of $p_{sec,p,2}^L$ can be obtained.
Finally substituting (\ref{psecvL1Planar}) and (\ref{psecvL2Planar}) into (\ref{psecvLPlanar}), the analytical result of $p_{sec,p}^L$ can be obtained.
\section{Proof of Theorem 6}
The secrecy  probability is
\begin{align}
&\mathrm{Pr}\left(\max_{e\in\left(\Phi_{e}+\Psi_{e}+\psi_e(l_o)\right)}\mathrm{SIR}_e\!\leq\! \beta\right)
\!\!=\!\!\mathbb{E}\left(
\underset{(\mathrm{I})}{\underbrace{\prod_{e\in\Psi_{e}}\mathrm{Pr}\left(\mathrm{SIR}_e\!\!\leq\!\!\beta\right)}}\times\right.\nonumber\\
&\left.
\underset{(\mathrm{II})}{\underbrace{\prod_{e\in\Phi_{e}}\mathrm{Pr}\left(\mathrm{SIR}_e\!\!\leq\!\!\beta\right)}}
\underset{(\mathrm{III})}{\underbrace{\prod_{e\in\psi_e(l_o)}\mathrm{Pr}\left(\mathrm{SIR}_e\!\!\leq\!\!\beta\right)}}\right).
\label{SececyOutage}
\end{align}

\begin{figure}[!t]
\centering
\subfigure[$\theta>\frac{\pi}{2}$] { \label{DistanceIllustration:a}
\includegraphics[width=0.46\columnwidth]{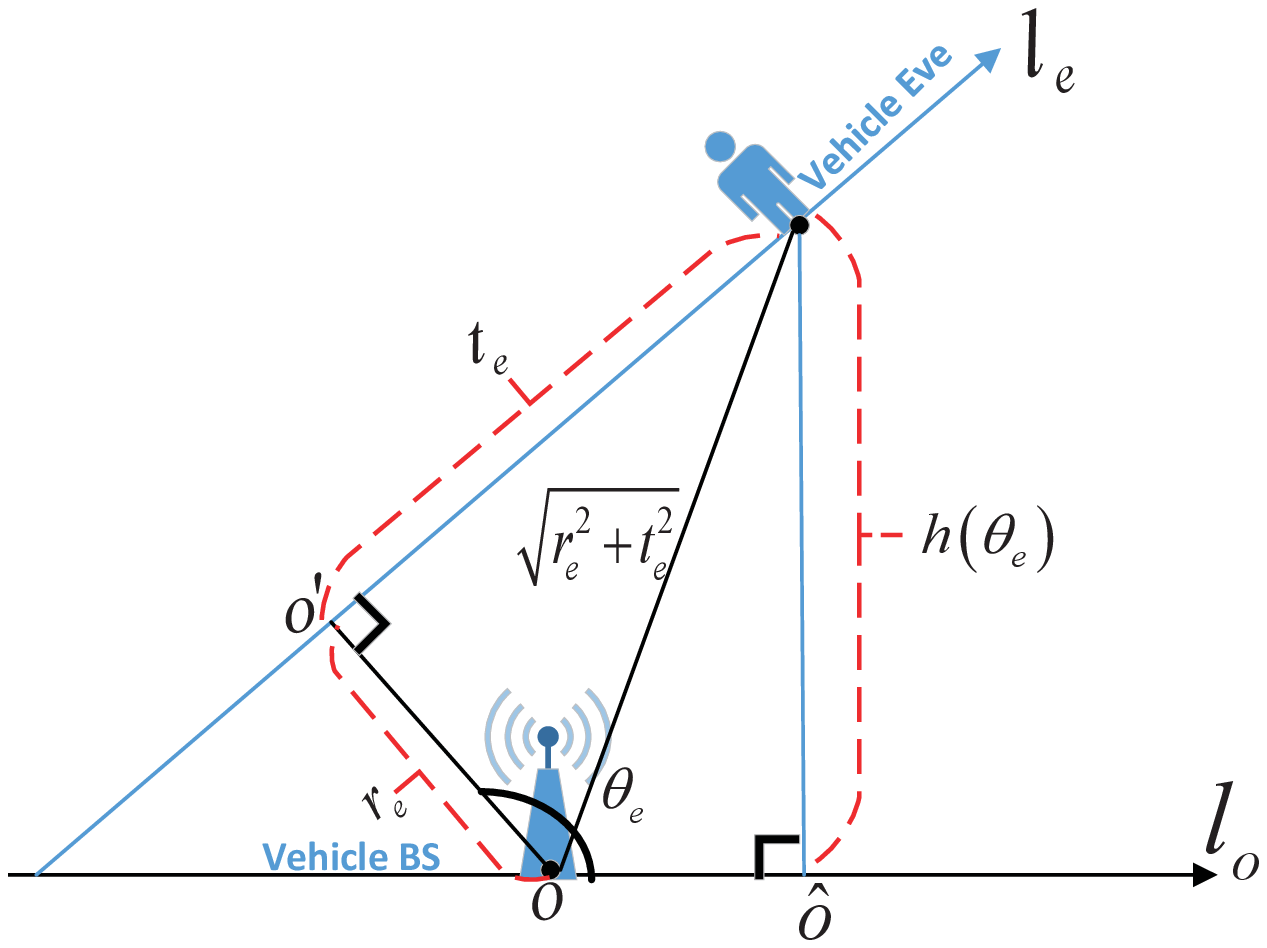}
}
\subfigure[$\theta<\frac{\pi}{2}$.] { \label{DistanceIllustration:b}
\includegraphics[width=0.46\columnwidth]{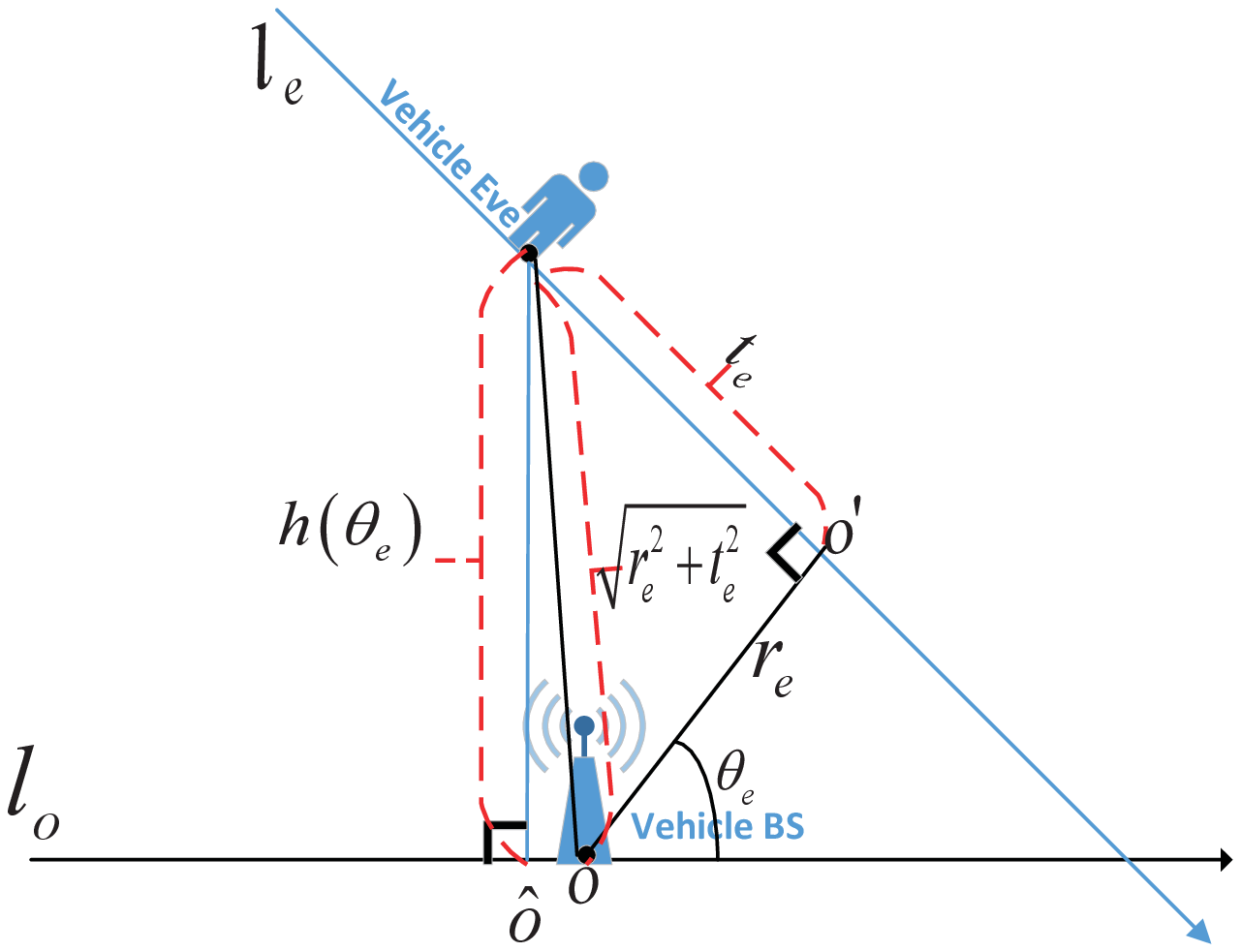}
}
\caption{Illustration of the relative location, where $o'$ is the projection of the origin $o$ onto the road $l_e$ and $\hat{o}$ is the projection of the vehicular Eve's location onto the road $l_o$. $h(\theta_e)$ is the perpendicular distance from the vehicular Eve to the road $l_o$.}
\label{DistanceIllustration}
\end{figure}

Following the procedures in (\ref{productIPlanar}), a lower bound on the expectation over the product $(\mathrm{I})$ in (\ref{SececyOutage}) can be derived as (\ref{productI}) at the top of the next page,
\begin{figure*}
\begin{align}
&\mathbb{E}\left(\prod_{e\in\Psi_{e}}\mathrm{Pr}\left(\mathrm{SIR}_e\leq\beta\right)\right)
{\gtrapprox}
\mathrm{exp}\left(-\frac{\lambda_l}{\pi}\int^{2\pi}_0d\theta_e\int^{+\infty}_0dr_e\mathrm{e}^{-u_e\int^{+\infty}_{-\infty}\mathrm{Pr}\left(\frac{\frac{\phi \left(N-1\right)}{1-\phi}|q_e|^2\left(r_e^2+t_e^2\right)^{-\frac{\alpha}{2}}}{
||\mathbf{g}_{oe}||_F^2\left(r_e^2+t_e^2\right)^{-\frac{\alpha}{2}}+\underset{{x\in\Delta_{b,v}}}{\sum}||\mathbf{g}_{xe}||_F^2D_{xe}^{-\alpha}}\geq\beta\right)dt_e}\right),
\label{productI}
\end{align}
\hrulefill
\end{figure*}
where $\Delta_{b,v}\triangleq{\Phi_b\cup\Psi_b\cup\psi_b(l_{e})\cup\psi_b(l_{o})}$.

Employing the PGFL of $\Phi_{e}$,
a lower bound on the expectation over the product $(\mathrm{II})$ in (\ref{SececyOutage}) is given by
\begin{align}
&\mathbb{E}\left(\prod_{e\in\Phi_{e}}\mathrm{Pr}\left(\mathrm{SIR}_e\leq\beta\right)\right)
{\gtrapprox}
\nonumber\\
&e^{-2\lambda_e \pi\int_0^{+\infty}\mathrm{Pr}\left(\frac{\frac{\phi(N-1)}{1-\phi}|q_e|^2r_e^{-{\alpha}}}{
 \frac{||\mathbf{g}_{oe}||_F^2}{r_e^{{\alpha}}}+\underset{{x\in\{\Phi_b+\Psi_b+\psi_b(l_o)\}}}{\sum}\frac{||\mathbf{g}_{xe}||_F^2}{D_{xe}^{\alpha}}}\geq\beta\right)r_edr_e}.
\label{productII}
\end{align}

Employing the PGFL of $\psi_e\left(l_o\right)$, a lower bound on the expectation over the product $(\mathrm{III})$ in (\ref{SececyOutage}) can be derived as
\begin{align}
&\mathbb{E}\left(\prod_{e\in\psi_e(l_o)}\mathrm{Pr}\left(\mathrm{SIR}_e\leq\beta\right)\right)
{\gtrapprox}
\nonumber\\
&e^{-2 u_e \int_0^{+\infty}\mathrm{Pr}\left(\frac{\frac{\phi(N-1)}{1-\phi}|q_e|^2t_e^{-{\alpha}}}{
\frac{ ||\mathbf{g}_{oe}||_F^2}{t_e^{{\alpha}}}+\underset{{x\in\{\Phi_b+\Psi_b+\psi_b(l_o)\}}}{\sum}\frac{||\mathbf{g}_{xe}||_F^2}{D_{xe}^{\alpha}}}\geq\beta\right)dt_e}
.\label{productIII}
\end{align}
Notice that under the condition that $e\in\psi_e(l_o)$,  the eavesdroppers are on the same road as the typical vehicular transmitter. Therefore, such eavesdroppers would suffer the interference from  the transmitters at $\Phi_b\cup\Psi_b\cup\psi_b(l_o)$.

For obtaining the analytical result of the  secrecy  probability, the ccdf of the SIR received by Eve should be derived first.

Defining $s\triangleq\frac{(\phi^{-1}-1)\left(r_e^2+t_e^2\right)^{\frac{\alpha}{2}}\beta}{N-1}$, and following the procedures in (\ref{ccdfSINRePlanar}), the ccdf in (\ref{productI}) can be derived as (\ref{ccdfSINRe}) at the top of the next page,
\begin{figure*}
\begin{align}
\mathrm{Pr}\left(\frac{\phi |q_e|^2\left(r_e^2+t_e^2\right)^{-\frac{\alpha}{2}}}{
\frac{\left(1-\phi\right)||\mathbf{g}_{oe}||_F^2}{\left(N-1\right)\left(r_e^2+t_e^2\right)^{\frac{\alpha}{2}}}+\underset{{x\in\{\Phi_b+\Psi_b+\psi_b(l_e)+\psi_b(l_o)\}}}{\sum}\frac{(1-u)||\mathbf{g}_{xe}||^2_FD_{xe}^{-\alpha}}{(N-1)}}\geq\beta\right)
{=}
\frac{\overset{(\mathrm{i})}{{\mathcal{L}^s_{\Phi_b}(s)}}
\overset{(\mathrm{ii})}{\mathcal{L}^s_{\Psi_b}(s)}
\overset{(\mathrm{iii})}{\mathcal{L}^s_{\psi_b(l_{r=0})}(s)}
\overset{(\mathrm{iv})}{\mathcal{L}^s_{\psi_b(l_{r=h(\theta_e)})}(s)}}{\left(1+\beta\frac{\phi^{-1}-1}{N-1}\right)^{N-1}},
\label{ccdfSINRe}
\end{align}
\hrulefill
\end{figure*}
where $h(\theta_e)$ is the perpendicular distance from the vehicular Eve to  $l_o$, as illustrated in Fig. \ref{DistanceIllustration}.

Applying Corollary 3, the analytical results of the Laplace transforms (i)-(iii) in (\ref{ccdfSINRe}) can be obtained.
For deriving the Laplace transform  (iv) in (\ref{ccdfSINRe}),
$h(\theta_e)$ should be derived firstly.
As illustrated in Fig. \ref{DistanceIllustration}, $h(\theta_e)=r_e\mathrm{sin}(\theta_e)-t_e\mathrm{cos}(\theta_e)$,  where  $t_e>0$ when the vehicular Eve is at the right of $o'$, otherwise, $t_e<0$. Employing Corollary 3, the analytical result of $\mathcal{L}^s_{\psi_b(l_{r=h(\theta_e)})}(s)$ can be obtained.
Then, substituting (\ref{ccdfSINRe}) into (\ref{productI}), the analytical result of $\mathbb{E}\left(\underset{{e\in\Psi_{e}}}{\prod}\mathrm{Pr}\left(\mathrm{SIR}_e\leq\beta\right)\right)
$ can be obtained.

With similar procedures, the analytical results of ${\underset{{e\in\Phi_{e}}}{\prod}\mathrm{Pr}\left(\mathrm{SIR}_e\leq\beta\right)}$ and ${\underset{{e\in\psi_e(l_o)}}{\prod}\mathrm{Pr}\left(\mathrm{SIR}_e\leq\beta\right)}$ can be obtained and the detailed derivations are omitted due to the space limitation.
Substituting (\ref{productI}), (\ref{productII}) and (\ref{productIII}) into (\ref{SececyOutage}), a lower bound on the secrecy probability can be obtained.

\ifCLASSOPTIONcaptionsoff
  \newpage
\fi



%
\bibliographystyle{IEEEtran}
\bibliography{SecurityV2X}

\end{document}